\documentclass[fleqn,usenatbib]{mnras}

\usepackage[T1]{fontenc}
\usepackage{ae,aecompl}
\usepackage{bm}
\usepackage{graphicx}    

\usepackage{amsmath}	
\usepackage{amssymb}	
\usepackage{txfonts}

\title[Polar planets are the most stable]{Polar planets around highly eccentric binaries are the most stable}

\author[Chen et al.]{Cheng Chen$^1$\thanks{Email: chenc21@unlv.nevada.edu},  Stephen H. Lubow$^2$ and Rebecca G. Martin$^1$
\\ $^1$Department of Physics and Astronomy,  University of Nevada, Las Vegas, 4505 South Maryland Parkway, Las Vegas, NV 89154, USA 
\\ $^{2}$Space Telescope Science Institute, 3700 San Martin Drive, Baltimore, MD 21218, USA\\}

\date{Accepted 15 April 2020. Received 31 March 2020; in original form 13 December 2019}

\pubyear{2020}
\begin{document}
\label{firstpage}
\pagerange{\pageref{firstpage}--\pageref{lastpage}}
\maketitle

\begin{abstract}
We study the orbital stability of a non-zero mass, close-in circular orbit planet around an eccentric orbit binary for various initial values of the binary eccentricity, binary mass fraction, planet mass, planet semi--major axis, and planet inclination by means of numerical simulations  that cover $5 \times 10^4$ binary orbits.
For small binary eccentricity, the stable orbits that extend closest to the binary (most stable orbits) are nearly retrograde and circulating. For high binary eccentricity, the most stable orbits are highly inclined and librate near the so-called generalised polar orbit  which is a stationary orbit that is fixed in the frame of the binary orbit. For more extreme mass ratio binaries, there is a greater variation in the size of the stability region (defined by initial orbital radius and inclination) with  planet mass and initial inclination, especially for low binary eccentricity.
For low binary eccentricity, inclined planet orbits may be unstable even at large orbital radii (separation $> 5 \,a_{\rm b}$).
The escape time for an unstable planet is generally shorter around an equal mass binary compared with an unequal mass binary. Our results have implications for circumbinary planet formation and  evolution and will be helpful for understanding future circumbinary planet observations.

\end{abstract}

\begin{keywords}
celestial mechanics -- planetary systems -- methods: analytical -- methods: numerical -- binaries: general
\end{keywords}

\section{Introduction}

Misaligned circumbinary discs are observed to be common in newborn binary systems \citep[e.g.,][]{Chiang2004,Winn2004,Capelo2012,Kennedy2012,Brinch2016,Kennedy2019}. This may be a result of the chaotic accretion process during star formation  \citep{Bateetal2003,bate2018} or stellar flybys \citep{clarke1993,Cuello2019b,Nealon2020}. Recent observations show that there is a break between alignment and misalignment of circumbinary discs with the central binary at orbital periods of about $30 \,\rm days $. About 68\% of short period binaries (period $<20\,\rm days $) have aligned disks (within 3$^{\circ}$), while those with longer orbital period show a larger range of inclinations and binary eccentricities \citep{Czekala2019}.  Since circumbinary planets form within a circumbinary disc, we expect their initial alignments to be similar to the disc. 

Misaligned discs should be more common around eccentric orbit binaries \citep{MartinandLubow2017}. Eccentric orbit
binaries occur at longer binary periods. Main sequence binaries are observed to be on circular orbits for periods
up to about 8 days due to tidal circularization \citep[e.g.,][]{Raghavan2010}. Binary eccentricities appear somewhat limited for binaries with periods of less than about 30 days, likely due to the effects of stellar tidal dissipation.
Of the ten circumbinary planets observed through transits by the Kepler mission, all of them are in orbits that are close to coplanar to their host binary orbit \citep{Welsh2012,Orosz2012a,Orosz2012b,Kostov2014,Welsh2015,Li2016,Kostov2016}. All the Kepler transit detected circumbinary planets orbit around a low eccentricity binary, except for Kepler-34b with a binary eccentricity of 0.52 \citep{Welsh2012,Kley2015}. However, the observed coplanarity may be a selection effect because these planets are found around binaries with smaller orbital periods and generally low binary eccentricity. In addition, the Kepler data may contain planets transiting non-eclipsing binaries and more inclined circumbinary planets may be found in the future \citep{martin2014}.

Binaries with longer orbital periods are expected to host planets with a wide range of inclinations (relative to the binary orbital plane) because their planet forming discs have a  wide range of inclinations. But planets around longer period binaries are harder to detect by transit methods because the transit probability is smaller and because the planet orbital period is longer.  Eclipse timing variations of the binary may be able to distinguish polar planets (those that are perpendicular to the binary orbital plane) from coplanar planets \citep{Zhang2019}. Furthermore, there is evidence based on binary eclipse timing variations for a highly misaligned circumbinary planet around the binary star KIC 5095269 \citep{Getley2017}. KIC 5095269b has an estimated mass of $7.70 \,M_{\rm J}$ and orbits a  binary with a moderate eccentricity of 0.26 \citep{Getley2017}. However, results from \cite{Borkovits2016} suggest that the binary eccentricity is lower $e_{\rm b}=0.05$ with a planet orbit inclination of $40^{\circ}$.

For a test (massless) particle that represents a low mass planet in orbit around a circular orbit binary, its angular momentum vector precesses around the angular momentum vector of the binary with constant tilt. Since the nodal precession rate does not change sign, the longitude of the ascending node fully circulates over 360$^{\circ}$. 
 The particle's orbit is  circulating and may be either prograde or retrograde relative to the binary, depending upon the initial particle inclination.
On the other hand, for a binary with non-zero eccentricity, a circumbinary test particle orbit may undergo libration if the test particle has a sufficiently large initial inclination. The longitude of the ascending node covers a limited range of angles, less than 360$^{\circ}$, and the nodal precession rate changes sign for a librating orbit.  In this case, the angular momentum vector of the planet precesses about the binary eccentricity vector  (semi-major axis) and undergoes tilt oscillations \citep{Verrier2009,Farago2010,Doolin2011,Naoz2017,deelia2019}. 

The minimum inclination required for a  planet to undergo libration decreases with increasing binary eccentricity and so a planet with even a small inclination may undergo libration in a highly eccentric binary. We define the stationary
orbit of a test particle to be the orbit which does not change secularly
in time in the inertial frame. That is, its orbit averaged tilt and longitude of ascending node are constant over long timescales. A particle orbit that is coplanar with a binary
has this property.
In addition, there is a stationary inclination state   in which the particle angular mementum vector lies along the binary semi-major axis (eccentricity vector). In this so-called polar aligned state, the orbital plane of the particle is perpendicular to the binary orbital plane. 

For a particle with nonzero mass, the binary orbit is no longer fixed. In this case, the stationary orbit, called the generalised polar orbit, refers to an orbit that is fixed in a frame comoves with the binary orbit.  That is, the orbit averaged tilt and longitude of ascending node of the generalised stationary orbit does not vary on long (secular) timescales in a frame defined by the orbit-averaged directions of the binary eccentricity and angular momentum. As the mass of the planet (particle) increases, the orbital inclination for the generalised polar aligned state decreases \citep{MartinandLubow2019,Chen20192}.

A circumbinary disc may also undergo  either circulation or libration depending on its initial inclination, binary eccentricity, and the disc angular momentum.  For a disc that is sufficiently warm and radially limited, it may undergo solid body precession \citep{LP1997}. Dissipation within the disc leads to alignment, either coplanar alignment to the binary orbital plane \citep{PT1995,LO2000,Nixonetal2011a,Nixonetal2012a,Facchinietal2013,Lodato2013,Foucart2013,Foucart2014} or polar alignment \citep{Aly2015,MartinandLubow2017,MartinandLubow2018b,Zanazzi2018,Franchini2019b,Smallwood2019b}. For an eccentric orbit binary, a disc undergoes tilt oscillations as it aligns, even if it is aligning to coplanar \citep{Smallwood2019}.
However, if the lifetime of the disc  is shorter than the disc alignment timescale of a planet-forming disc, then planets may form in a misaligned disc that is neither coplanar nor polar. Thus, a giant planet may have a misaligned orbit around a binary system. The planet and the disc may not remain coplanar with each other once the planet opens a gap in the disc  \citep{Picogna2015,Lubow2016,Martin2016,Franchini2019c}.

In this work, we study the orbital stability of misaligned close-in planets around eccentric binaries for different binary and planet parameters. \citet{Holman1999}, \cite{Sutherland2016}, and \cite{Quarles2018} considered the orbital evolution and stability of a test particle on an initially coplanar circular orbit about an eccentric binary. \citet{Doolin2011} extended these test particle results to the noncoplanar case \citep[see also][]{Quarles2016,Hong2019}. In \citet{Chen20192} we further extended these studies to investigate the evolution of an inclined circumbinary planet with mass on an initially circular orbit. The orbital behaviour of the planet is in good agreement with the analytic results of \cite{MartinandLubow2019} that is based on the quadrupole approximation for the binary potential \citep{Farago2010}.   The agreement breaks down for close-in planets where the quadrupole approximation is less accurate. In the octupole approximation, these orbits have only been studied for test particles and not planets with mass \citep{Naoz2017,Vinson2018}. 

We found that there is a generalised polar orbit whose inclination angle depends on the planet mass. This orbit is stationary in a frame that precesses with the binary. Recently, \cite{Cuello2019} and \cite{Giuppone2019} included an analysis of the orbital dynamics of massive P-type planets around eccentric binaries. They considered up to Jupiter mass planets with an orbital inclination of $90^\circ$. This configuration does not provide a generalised polar (stationary) orbit, as is consistent with their findings.  We describe the initial conditions for the three body simulations and  show the orbital stability maps in Section~\ref{results}. In Section~\ref{conclusion}, we  discuss the implications of our results and draw our conclusions.

\section{Three body simulations} 
\label{results}

In this Section we first describe the simulation set--up and the parameter space that we explore. We then show the stability maps that descibe the stable circumbinary orbits. Finally, we consider the timescale on which the unstable orbits become unstable. 

\subsection{Simulation set--up and parameter space explored}

\begin{table}
\centering
\caption{The initial conditions for the planet orbits in the three body simulations. The planet is initially in a circular Keplerian orbit about the centre of mass of the binary. The initial planet eccentricity is $e=0$, the argument of periapsis is $\omega=0$ and  the longitude of the ascending node is $\phi=90^\circ$. Column~1 shows the name of the orbital parameter. Column~2 shows the minimum value in the phase space. Column~3 shows the maximum value in the phase space. Column~4 shows the spacing in the phase space.  }

\begin{tabular}{cccc} 
\hline
Orbital element & Minimum value & Maximum value & $\Delta$ \\
\hline
\hline
$a$ & 1.5 $a_{\rm b}$ & 6 $a_{\rm b}$ & 0.05 $a_{\rm b}$\\
$i$ & 0 & $\pi$ & $\pi$/80\\
$\nu$ & 0 & 5$\pi$/3 &$\pi$/3\\
\hline
\label{table2}
\end{tabular}
\end{table}

\begin{table}
\centering
\caption{Parameters of the simulations. The first column contains the name of the Model, the second and third columns indicate the binary  eccentricity and mass fraction. The fourth column represents the mass of the planet in units of $m_{\rm b}$ }
\begin{tabular}{cccc} 
\hline
\textbf{Model} & $e_{\rm b}$ & $f_{\rm b}$ & $m_{\rm p}$ ($m_{\rm b}$) \\
\hline
\hline
A1 &  0.2  & 0.5 & 0.001 \\
A2 &  0.2  & 0.1 & 0.001 \\
A3 &  0.2  & 0.5 & 0.005  \\
A4 &  0.2  & 0.1 & 0.005  \\
A5 &  0.2  & 0.5 & 0.01  \\
A6 &  0.2  & 0.1 & 0.01  \\
\hline
B1 &  0.5  & 0.5 & 0.001  \\
B2 &  0.5  & 0.1 & 0.001  \\
B3 &  0.5  & 0.5 & 0.005  \\
B4 &  0.5  & 0.1 & 0.005  \\
B5 &  0.5  & 0.5 & 0.01   \\
B6 &  0.5  & 0.1 & 0.01  \\
\hline
C1 &  0.8  & 0.5 & 0.001  \\
C2 &  0.8  & 0.1 & 0.001  \\
C3 &  0.8  & 0.5 & 0.005  \\
C4 &  0.8  & 0.1 & 0.005  \\
C5 &  0.8  & 0.5 & 0.01  \\
C6 &  0.8  & 0.1 & 0.01  \\
\hline
\label{table1}
\end{tabular}
\end{table}

To study the stability of a third body orbiting around an eccentric binary star system, we use a {\sc whfast} integrator which is a second order symplectic Wisdom Holman integrator with 11th order symplectic correctors in the ${\sc n}$-body simulation package, {\sc rebound} \citep{Rein2015b}. We solve the gravitational equations for the three bodies in the frame of the centre of mass of the three body system for which the central binary has components of mass $m_1$ and $m_2$ with total mass $m_{\rm b}=m_1+m_2$. The mass fraction of the binary is defined with $f_{\rm b}=m_2/m_{\rm b}$. The separation of the binary is  $a_{\rm b}$, the eccentricity of the binary is $e_{\rm b}$ and the orbital period of the binary is $T_{\rm b}$. 

The planet is initially in a Keplerian orbit around the centre of mass of the binary and has mass $m_{\rm p}$. Its orbit is defined by six orbital elements: the semi-major axis $a$, inclination  $i$ relative to the binary orbital plane, eccentricity $e$, longitude of the ascending node $\phi$ measured from the binary semi--major axis, argument of periapsis $\omega$, and true anomaly $\nu$. The orbit of the planet is initially circular and so initially $e=0$ and  $\omega=0$. We choose $\phi=90^{\circ}$ or $\phi=0^{\circ}$  initially in our suites of simulations. In Table~\ref{table2}, we describe the sampling of the phase space for how we vary  $a$, $i$ and $\nu$. Note that the binary orbit is not fixed since the binary feels the gravity of the massive third body. 

In this work we are interested in the stability of circumbinary planet orbits that are close to the binary. 
We consider orbits for which the initial semi--major axis $1.5 a_{\rm b} \le a \le 6 a_{\rm b}$ over times of $5\times 10^4 T_{\rm b}$ and over the full range of initial inclinations. We define the orbit as unstable once at least one of three criteria are met.  Instability occurs first, if the eccentricity of the planet becomes large $e > 1.0$ so that the planet is not bound to the binary; second, if the semi--major axis of the planet increases significantly,  $a > 10\, a_{\rm b}$; or third, if the semi-major axis of the planet becomes very small, $a <  a_{\rm b}$ \citep[see also, for example][]{Quarles2019}.  We define the escape time to be the time at which one of these criteria is satisfied. 
If the planet does not meet any of the three criteria for instability within the time of $5\times 10^4 T_{\rm b}$, then we classify it as stable.

In order to classify the orbital motion of the planet, we work in a frame relative to the instantaneous eccentricity and angular momentum vectors of the binary ($\bm{e_{\rm b}}$ and $\bm{l_{\rm b}}$).  
The frame  has the three axes along unit vectors
$\bm{\hat{e}_{\rm b}}$, $\bm{\hat{l}_{\rm b}}\times \bm{\hat{e}_{\rm b}}$, and $\bm{\hat{l}_{\rm b}}$. For
planet angular momentum $\bm{l_{\rm p}}$,  the inclination of the planet's orbital plane relative to the binary orbital plane is given by
\begin{equation}
i =\cos^{-1}(\bm{\hat l_{\rm b}}\cdot \bm{\hat l_{\rm p}}),
\end{equation}
where $\bm{\hat l_{\rm b}}$ is a unit vector in the direction of the
angular momentum of the binary and $\bm{\hat l_{\rm p}}$ is a unit
vector in the direction of the angular momentum of the
planet. The inclination of the binary relative to the total angular momentum $\bm{l}$ is
\begin{equation}
\i_{\rm b} =\cos^{-1}(\bm{\hat l}\cdot \bm{\hat l_{\rm b}}),
\end{equation}
where $\bm{\hat l}$ is a unit
vector in the direction of the total angular momentum ($\bm{l}=\bm{l}_{\rm b}+\bm{l}_{\rm p}$). Similarly, the phase angle (longitude of ascending node) of the particle in the
same frame of reference is given by
\begin{equation}
\phi=\tan^{-1}\left(\frac{ \bm{\hat{l}}_{\rm p}\cdot (\bm{\hat{l}}_{\rm b}\times \bm{\hat{e}}_{\rm b})}{ \bm{\hat{l}}_{\rm p}\cdot \bm{\hat{e}}_{\rm b}}  \right) + 90^{\circ}.
\end{equation}
(Note that this equation corrects equation 3 in \citet{Chen20192} that contains some typos.)

There are four types of orbits in our stability maps. These are shown in the $i\cos \phi$--$i \sin \phi$ phase space in Figure 1 in \cite{Chen20192}. There are two types of circulating orbits, meaning the planet angular momentum precesses about the binary angular momentum vector. The green lines represent prograde orbits that display retrograde (clockwise) precession, $ d\phi/dt < 0$, and the blue lines represent retrograde orbits that display prograde (counterclockwise) precession, $ d\phi/dt > 0$. The librating orbits, meaning  those where the planet angular momentum vector precesses about the direction of angular momentum of the stationary orbit, are shown in the cyan and red lines. The stationary inclination for a test particle is $90^\circ$, but the generalised stationary state for a planet with mass is at lower inclination. The inclination at the centre of the librating orbits is the stationary inclination, $i_{\rm s}$. The red orbits have initial inclination $i<i_{\rm s}$, while the cyan orbits have initially $i>i_{\rm s}$. These librating orbits display prograde precession in $\phi$. The centre is at $i=i_{\rm s} \approx 90^\circ$ for a low mass planet case and it is < $90^\circ$ for a high mass planet case \citep[see equation 5 in][]{Chen20192}.

Table~\ref{table1} lists the values of the initial binary eccentricity, binary mass fraction, and planet mass for all of the simulations that we consider. We explore three binary eccentricities, $e_{\rm b}=0.2$, 0.5, and 0.8, two binary mass fractions $f_{\rm b}=0.5$ and 0.1, and three different planet masses, $m_{\rm p}=0.001\,m_{\rm b}$,  $0.005\,m_{\rm b}$, and $0.01\,m_{\rm b}$.

In the case with initial $\phi=90^\circ$ there are always librating orbits. This can be seen for example from Fig.~1 in \cite{Chen20192}.  For test particles all possible circulating and librating orbits with fixed binary eccentricity are covered by sampling along the vertical line in the $i\cos \phi - i \sin \phi$, corresponding to $\phi=90^\circ$. In the non-zero mass planet case,  the vertical line does not sample all possible orbits at fixed initial binary eccentricity because the binary eccentricity varies in time. For example, for the case with initial $\phi=90^\circ$, two planet orbits with different initial inclinations but the same initial binary eccentricity  have different binary eccentricity by the time they reach $\phi=0^\circ$. We are considering stability maps where the initial binary eccentricity is the same across all orbits so it is necessary to consider orbits at different initial phase angles. For each model, we consider two initial values for the nodal phase angle, $\phi=90^\circ$ and $\phi=0^\circ$.  The case of initial $\phi=90^\circ$ is most favourable for producing librating orbits, while the case of initial $\phi=0^\circ$ does not produce any librating orbits.

\subsection{Stability maps for initial $\phi=90^\circ$}

\begin{figure*}
    \centering
    \includegraphics[width=5.8cm]{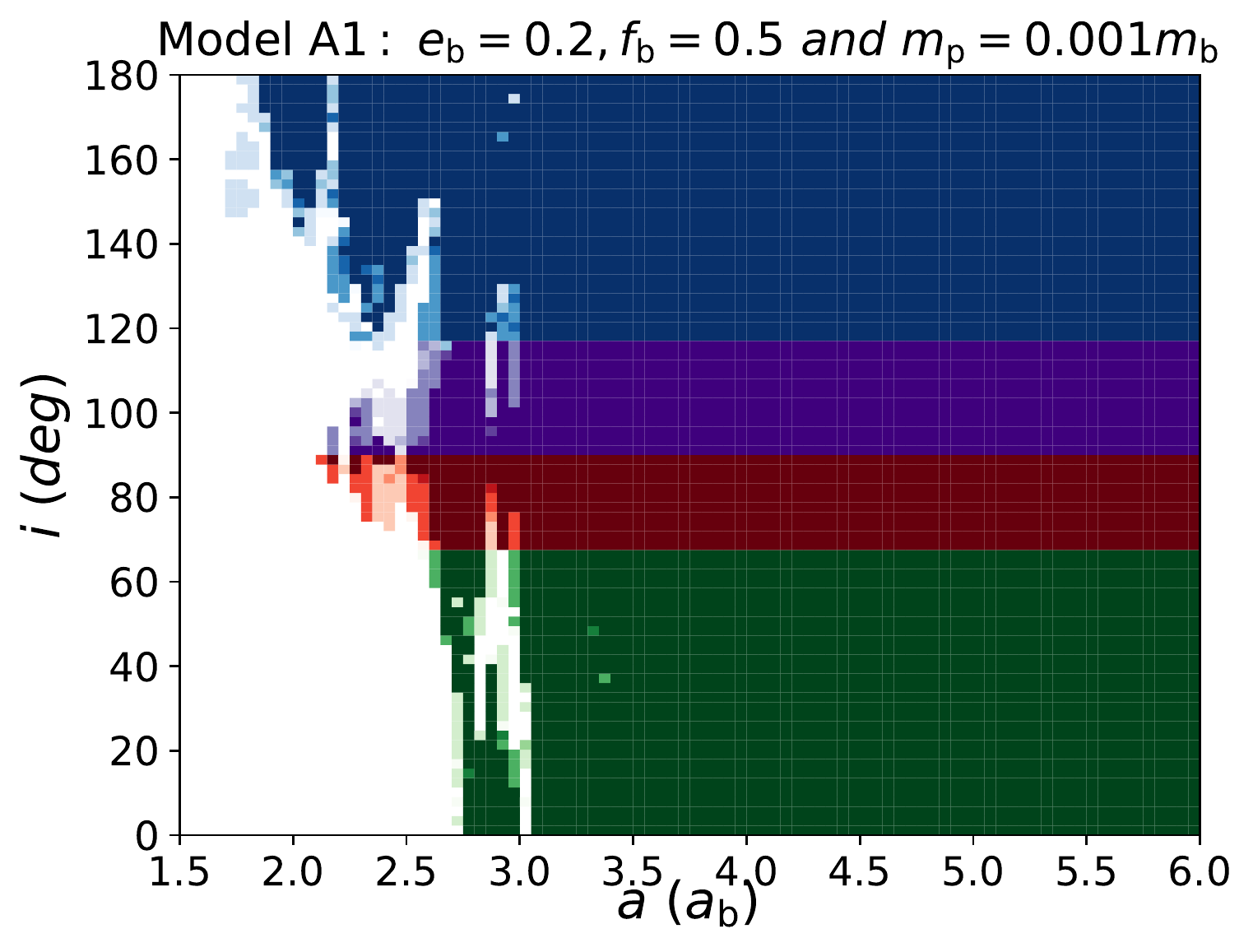}
    \includegraphics[width=5.8cm]{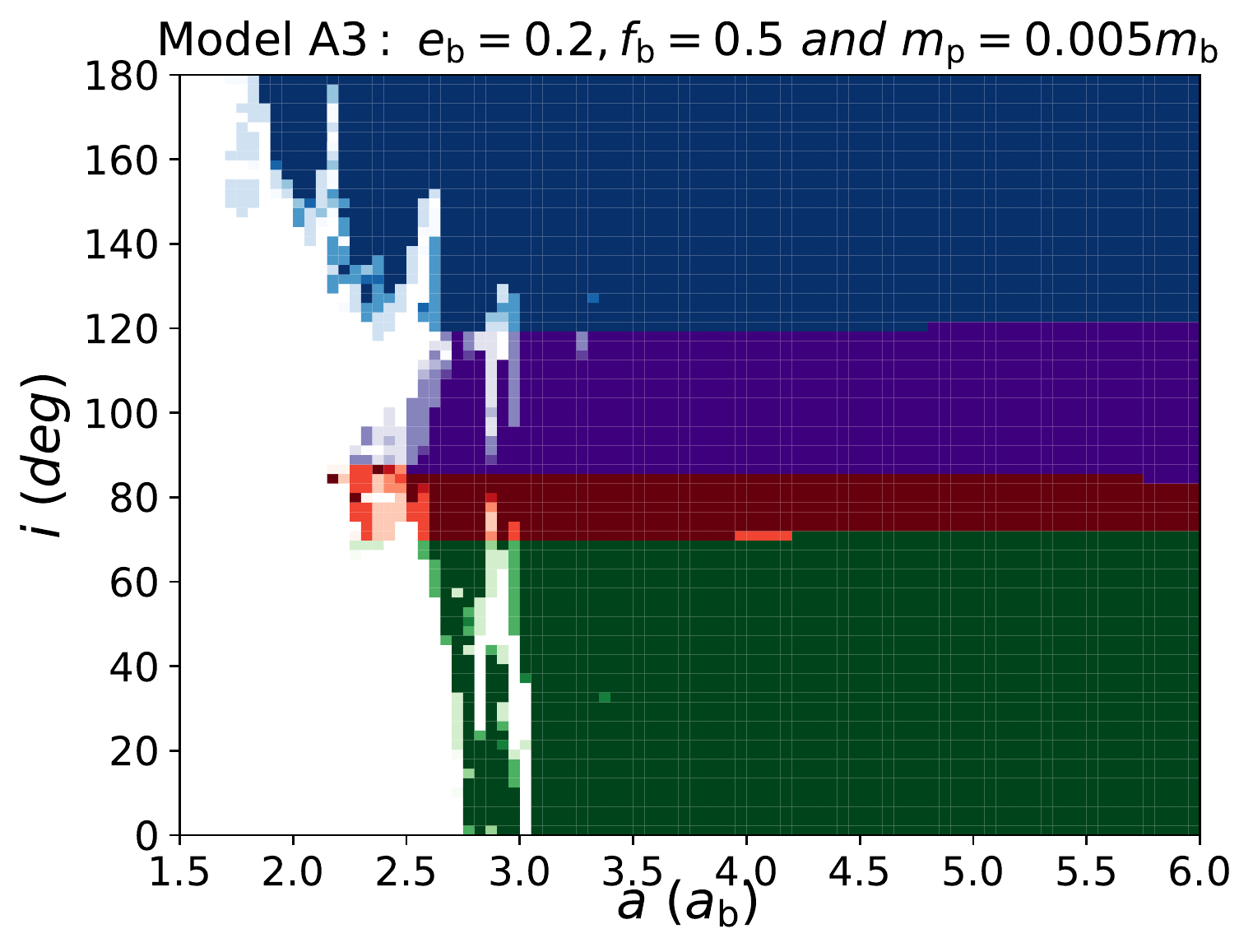}
    \includegraphics[width=5.8cm]{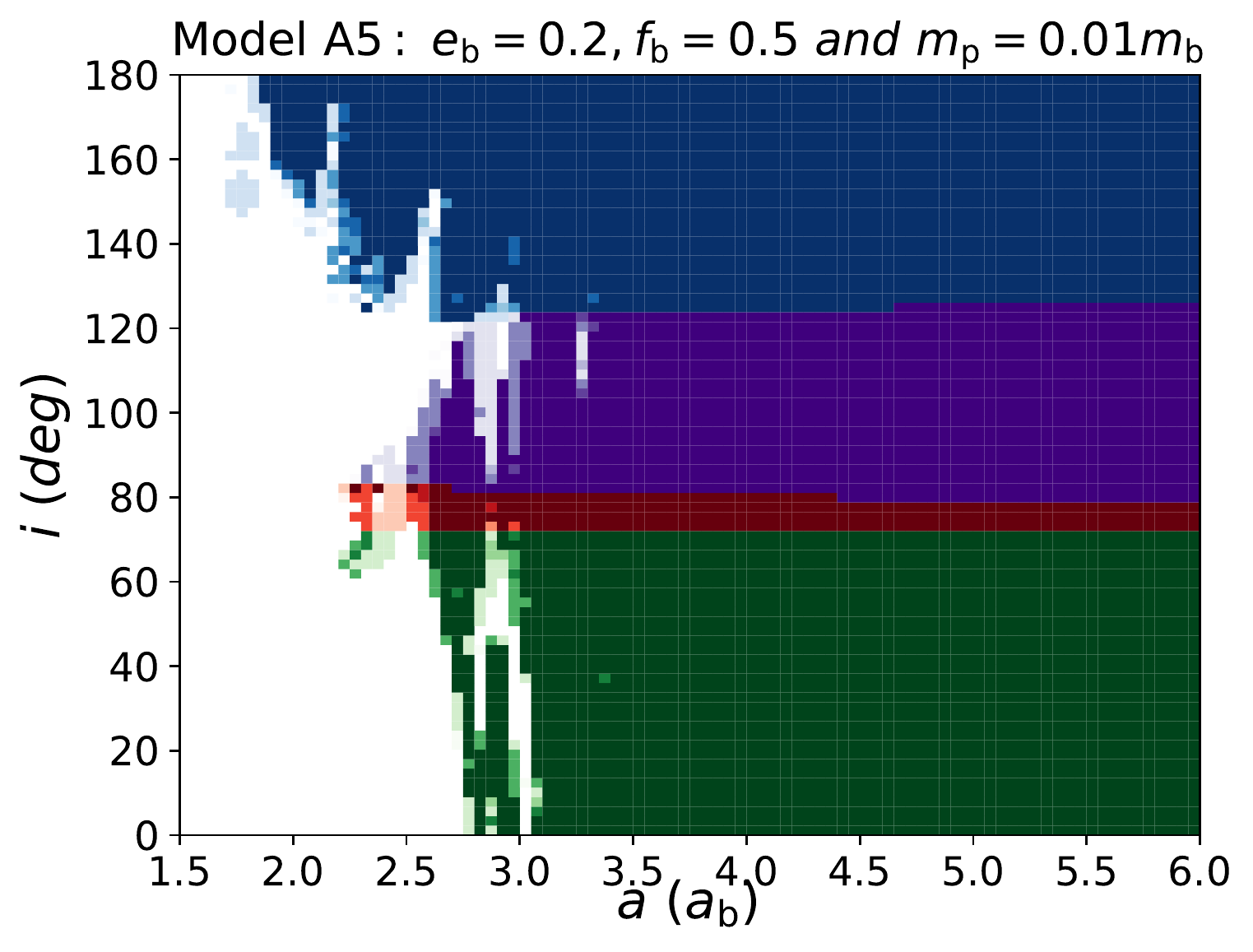}
    \includegraphics[width=5.8cm]{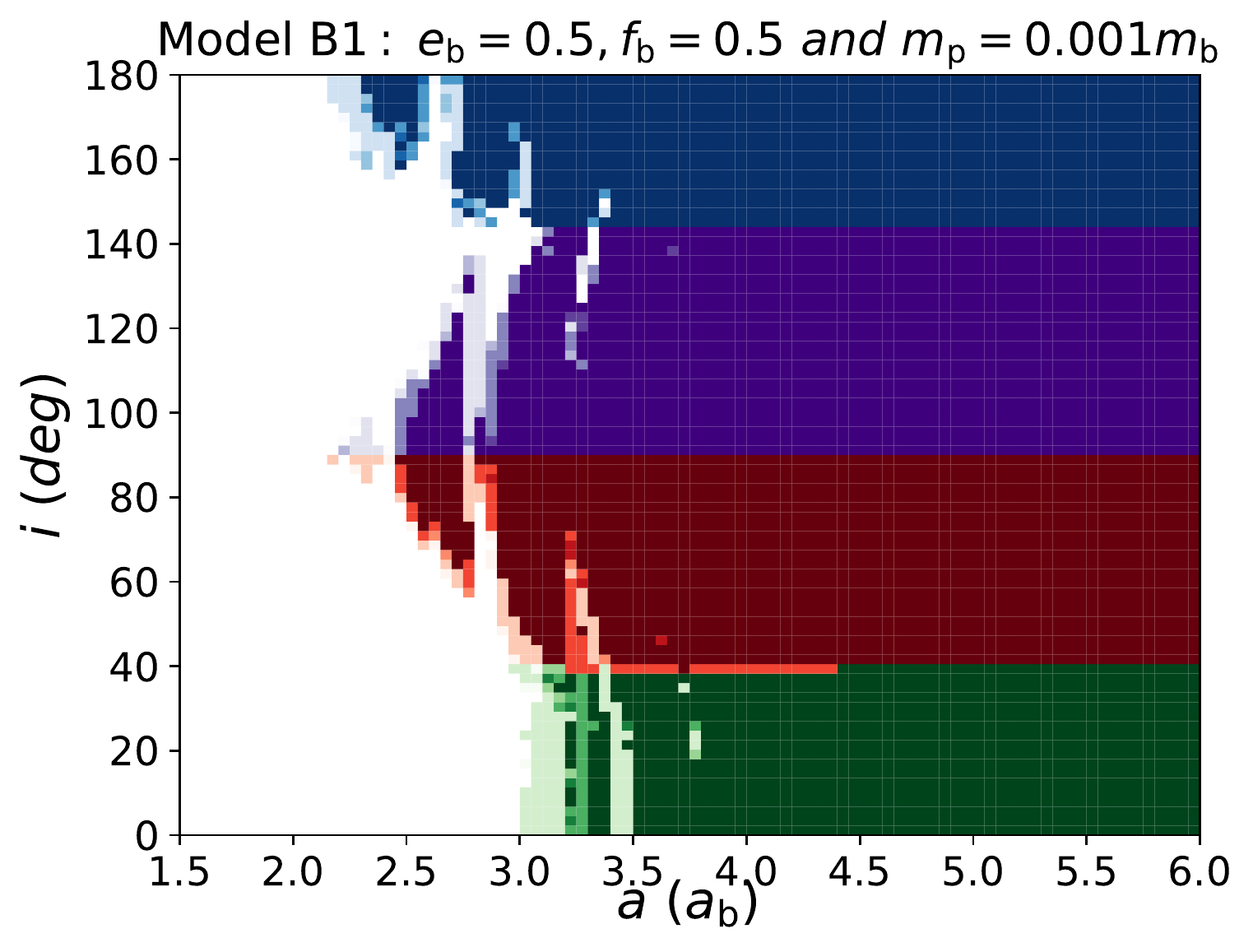}
    \includegraphics[width=5.8cm]{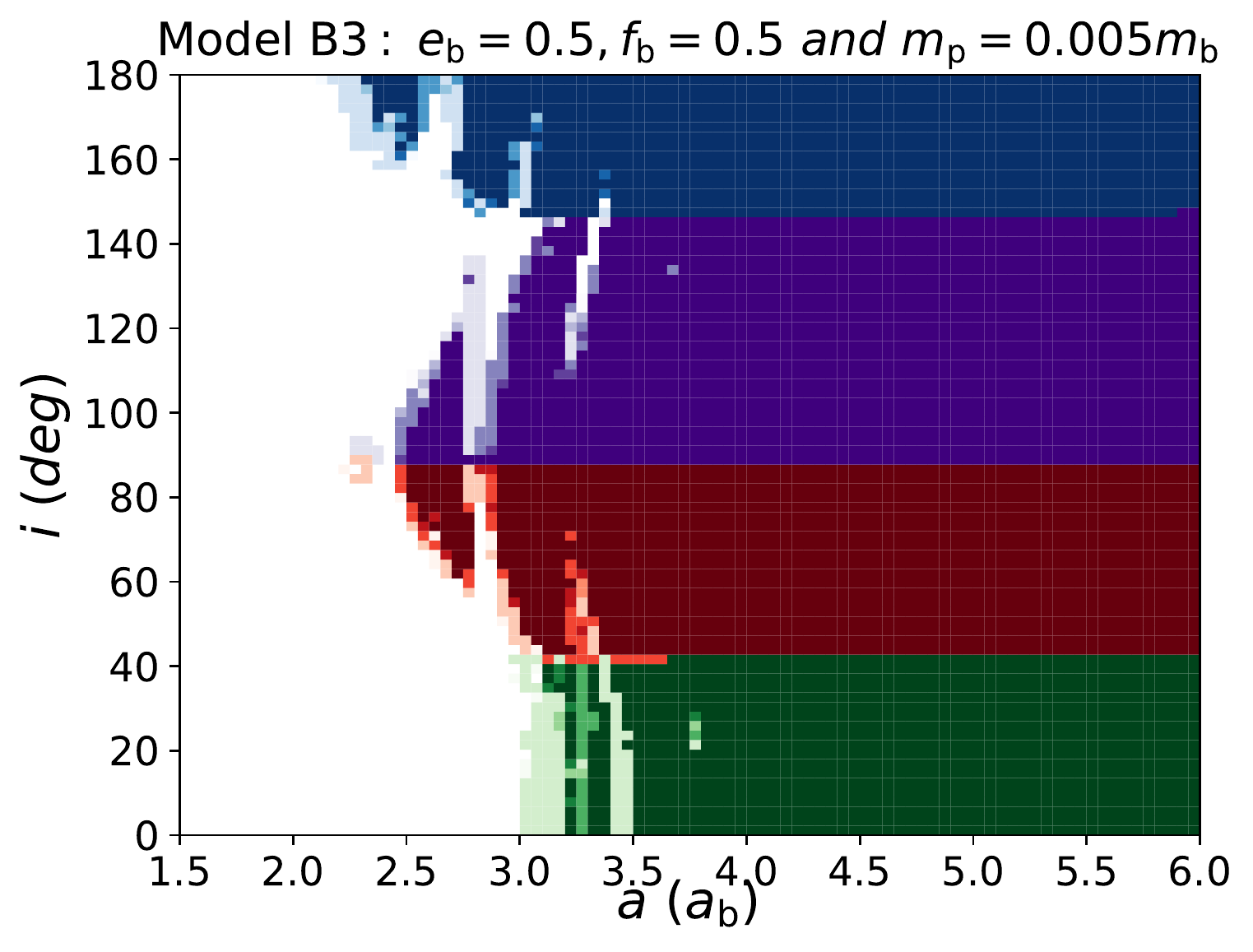}
    \includegraphics[width=5.8cm]{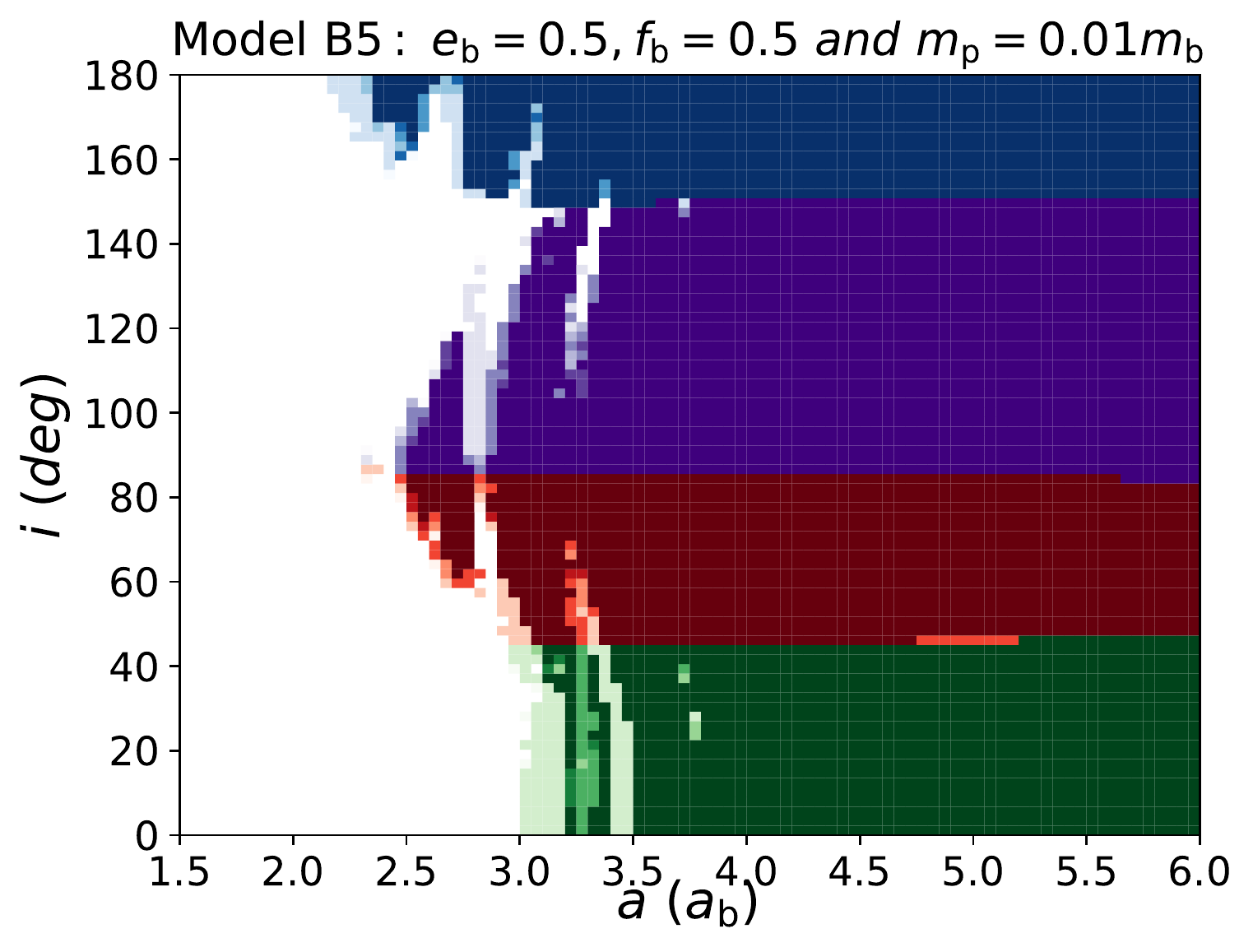}
    \includegraphics[width=5.8cm]{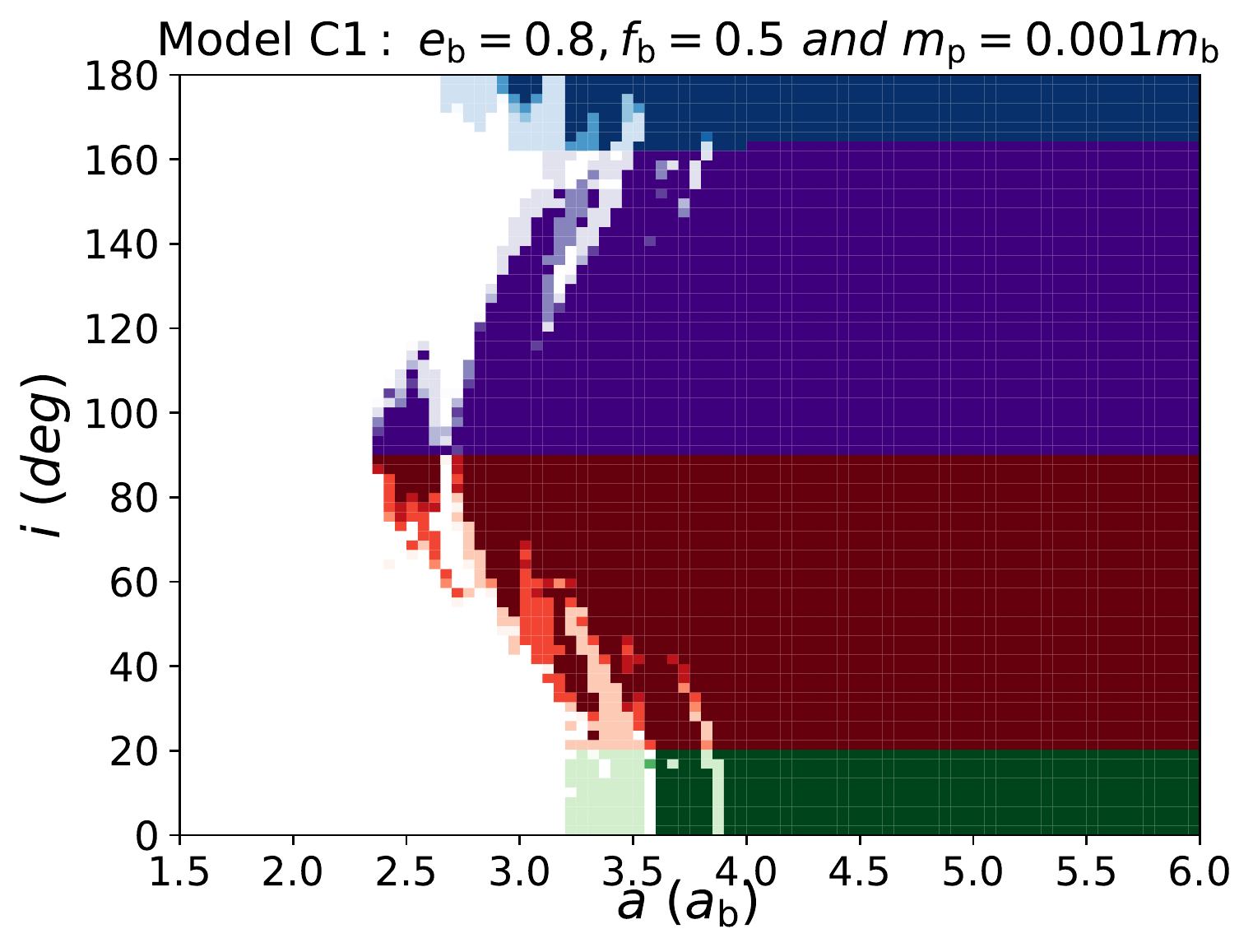}
    \includegraphics[width=5.8cm]{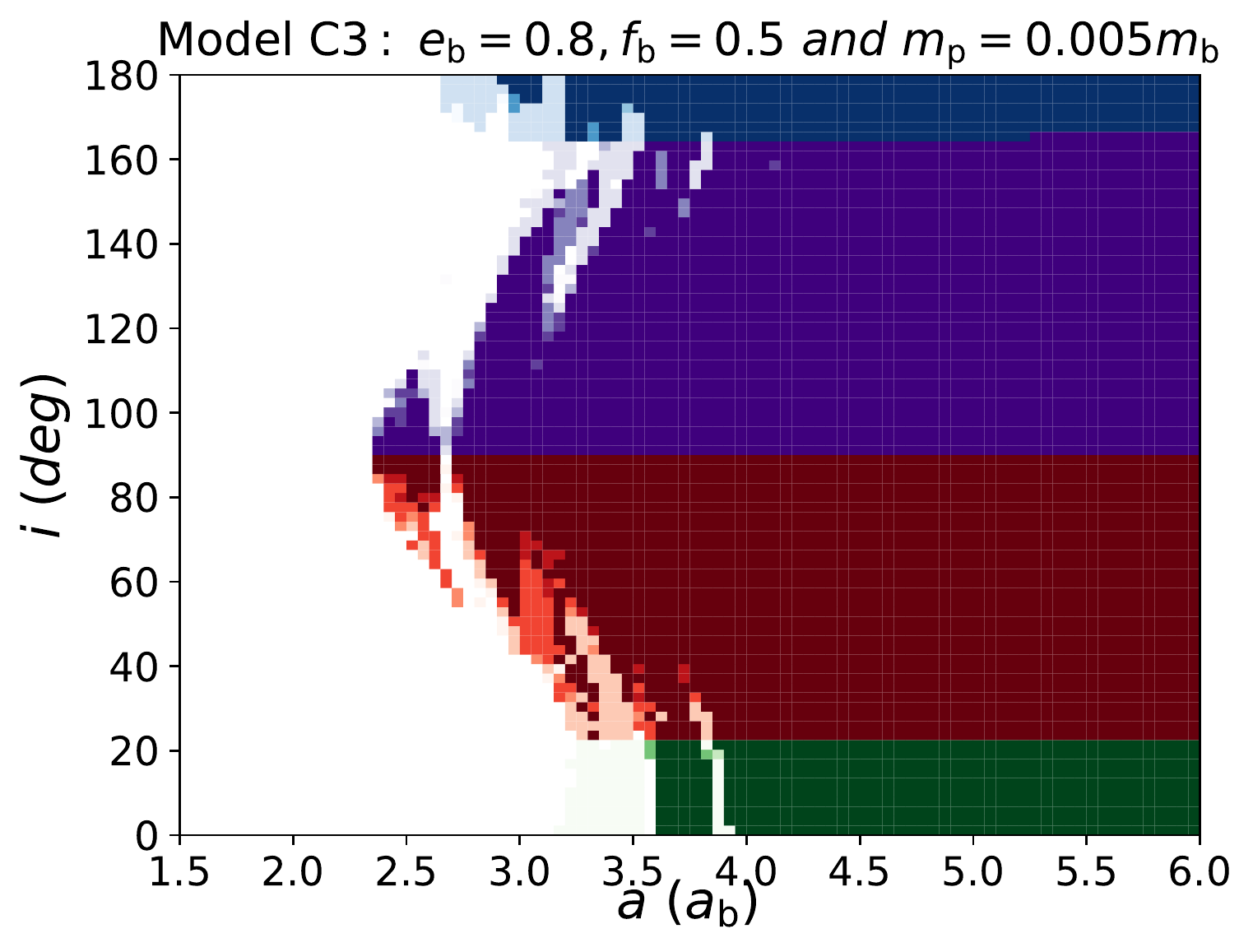}
    \includegraphics[width=5.8cm]{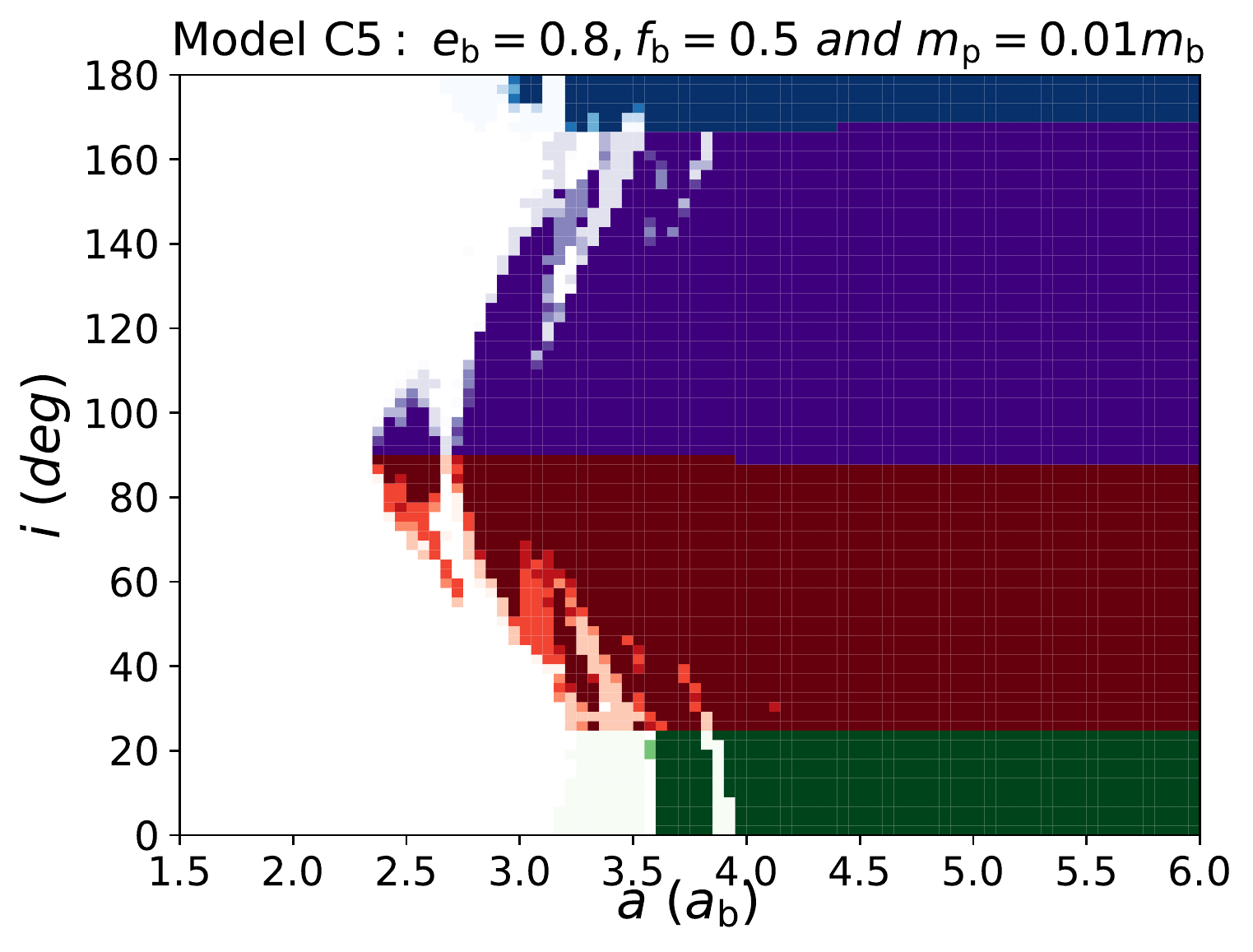}
    \caption{Orbital stability as a function of initial planet separation $a$ and initial  inclination $i$ for initial $\phi=90^\circ$. The binary is equal mass, $f_{\rm b}=0.5$, and the initial binary eccentricity is $e_{\rm b}$ = 0.2 (first row), 0.5 (second row), and 0.8 (third row). The third body has mass  $m_{\rm p}=10^{-3}\,m_{\rm b}$ (first column), $5\times 10^{-3}\,m_{\rm b}$ (second column), and $10^{-2}\,m_{\rm b}$ (third column). The green and blue pixels represent prograde and retrograde circulating orbits, respectively. The red pixels represent librating orbits with initial inclination $i<i_{\rm s}$ while the purple pixels represent librating orbits with initial inclination $i>i_{\rm s}$. Each pixel represents  simulations for six different values of the true anomaly and the  darker the colour, the more stable orbits. The white pixels represent unstable orbits. }
    \label{fig:map1}
\end{figure*}

\begin{figure*}
    \centering
    \includegraphics[width=5.8cm]{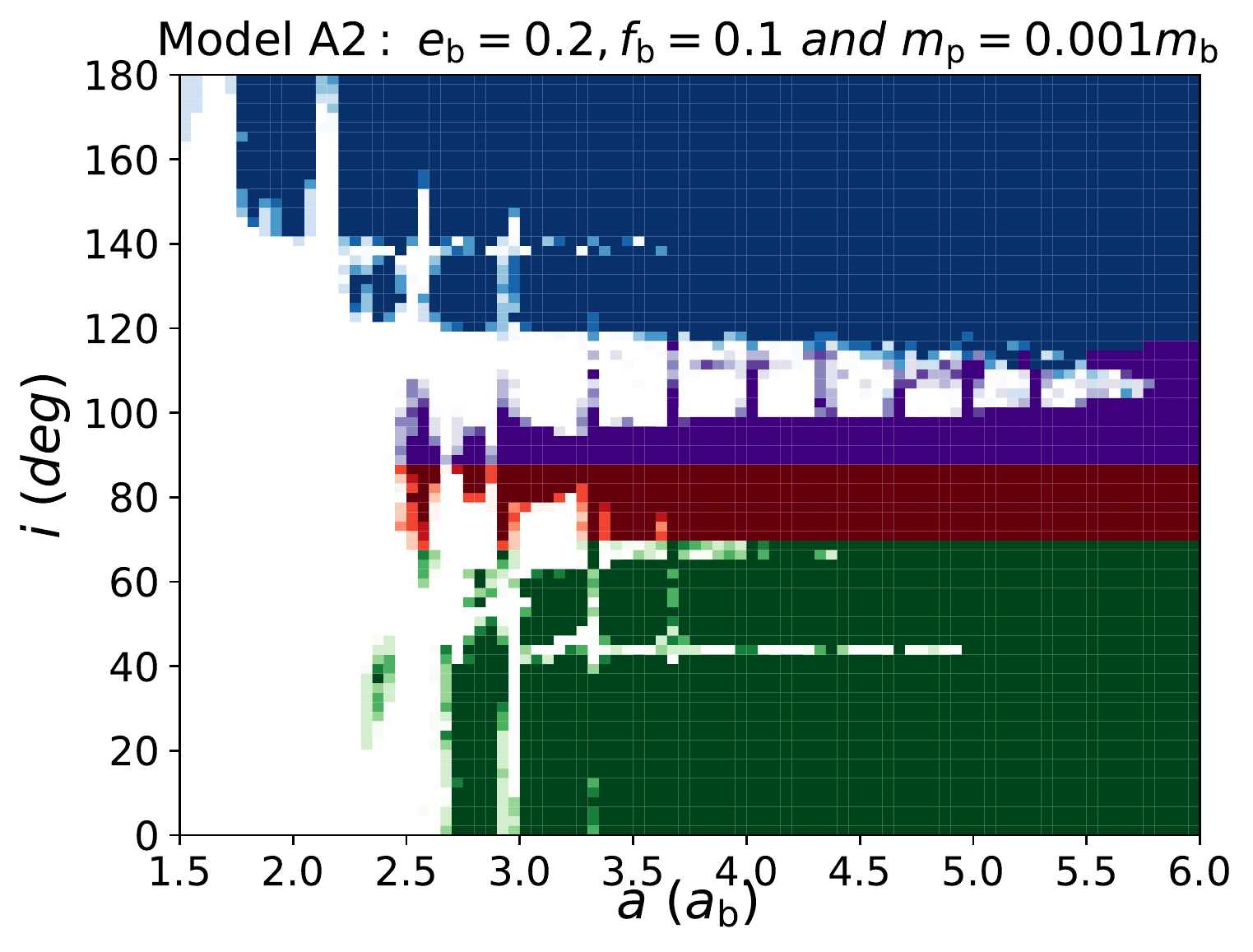}
    \includegraphics[width=5.8cm]{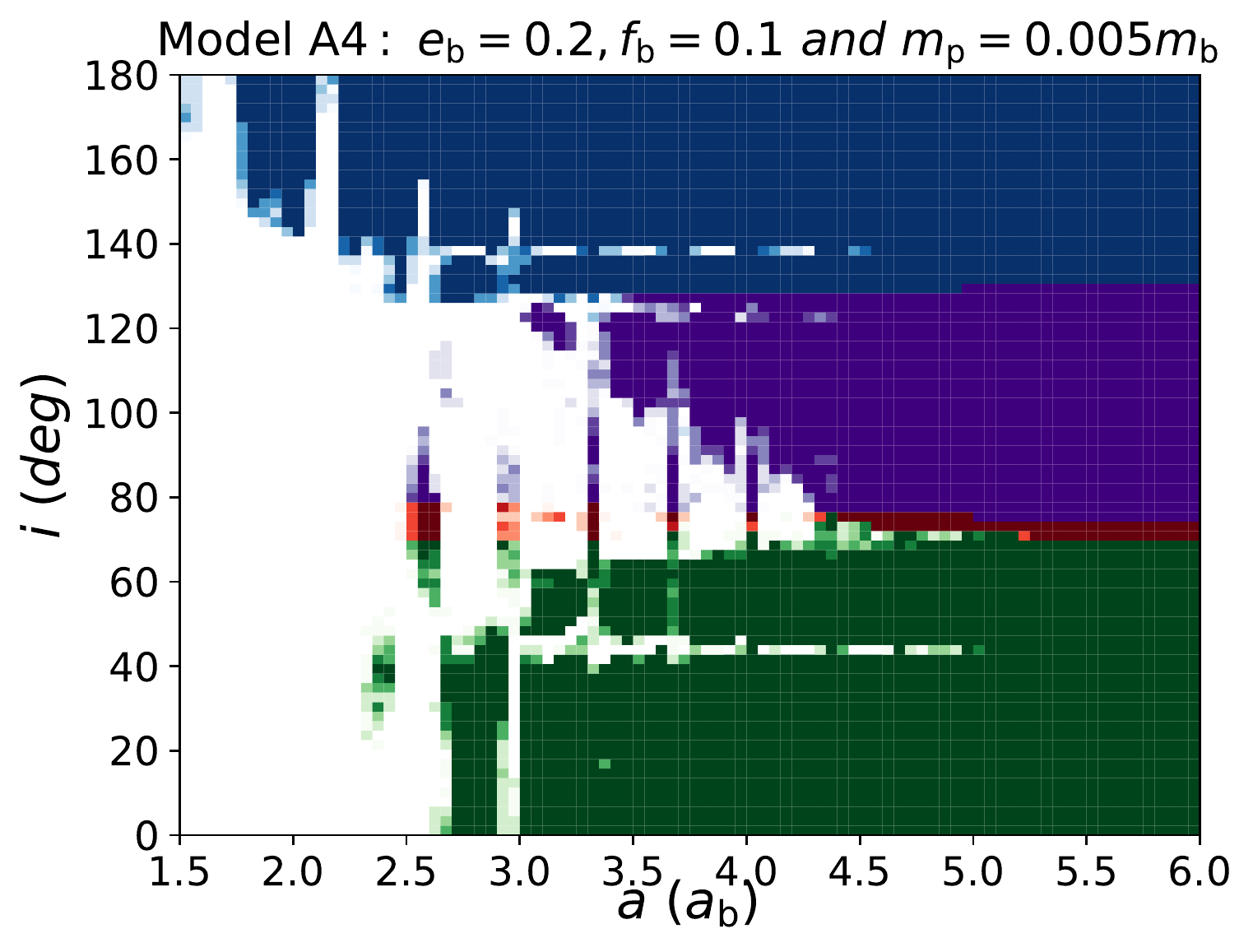}
    \includegraphics[width=5.8cm]{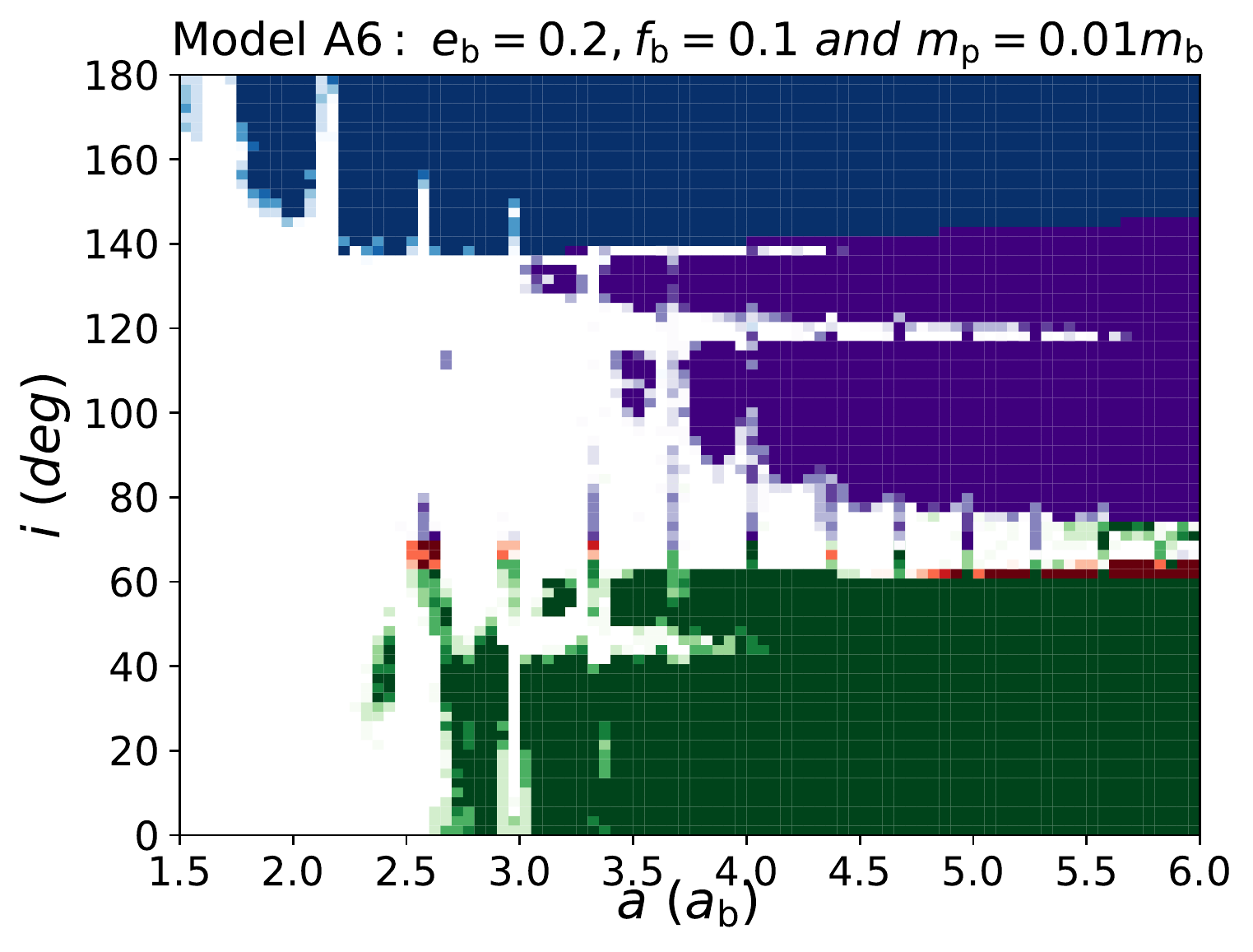}
    \includegraphics[width=5.8cm]{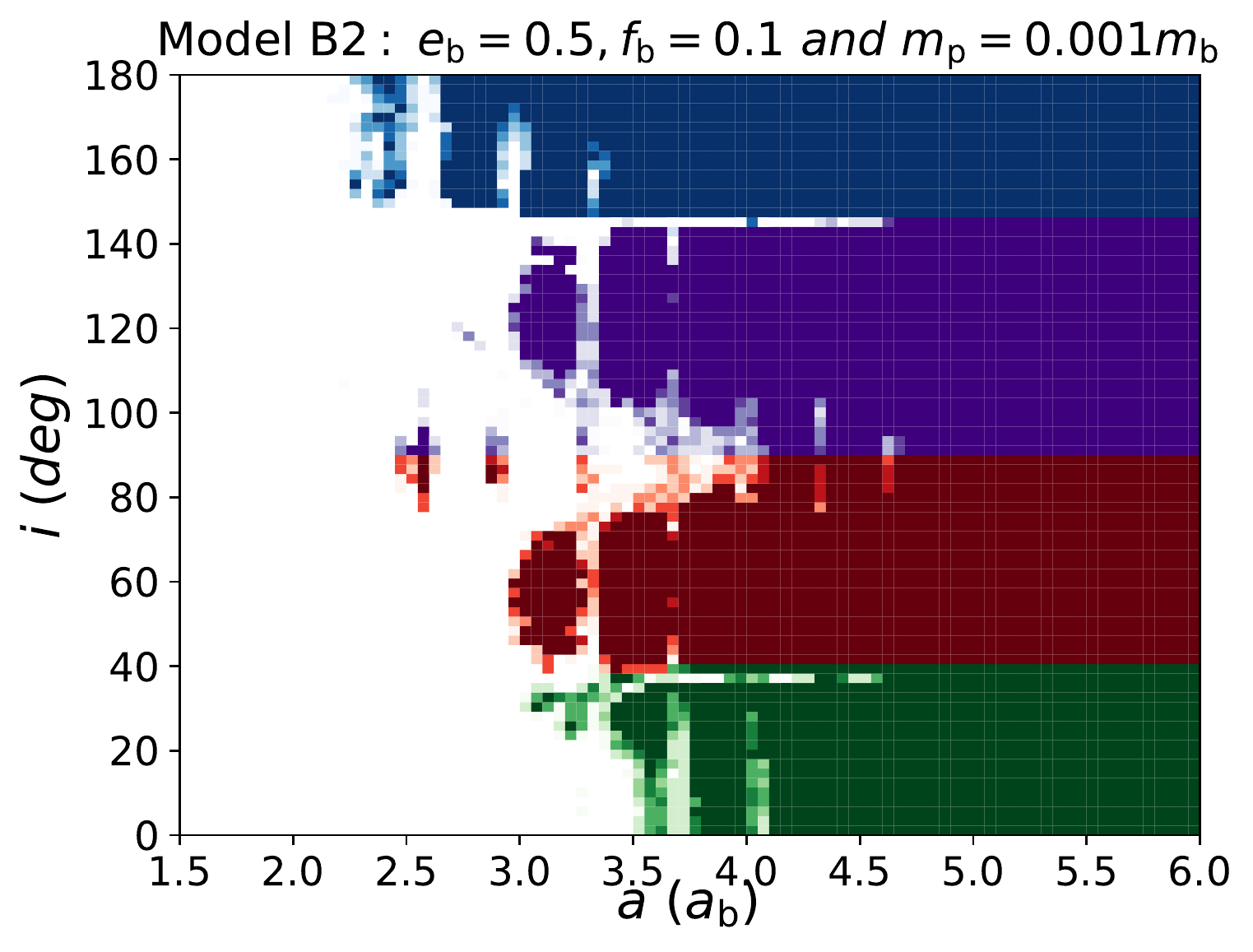}
    \includegraphics[width=5.8cm]{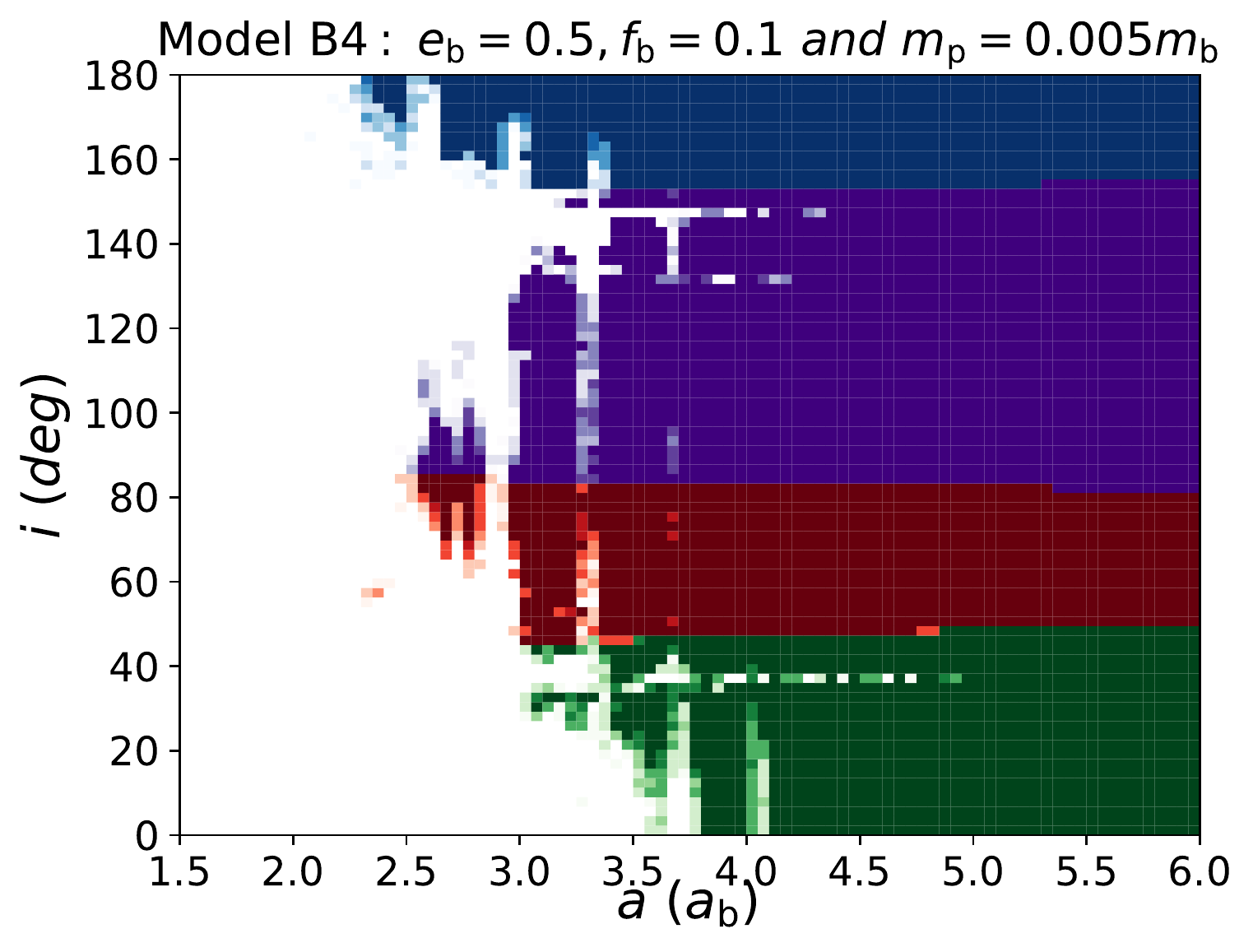}
    \includegraphics[width=5.8cm]{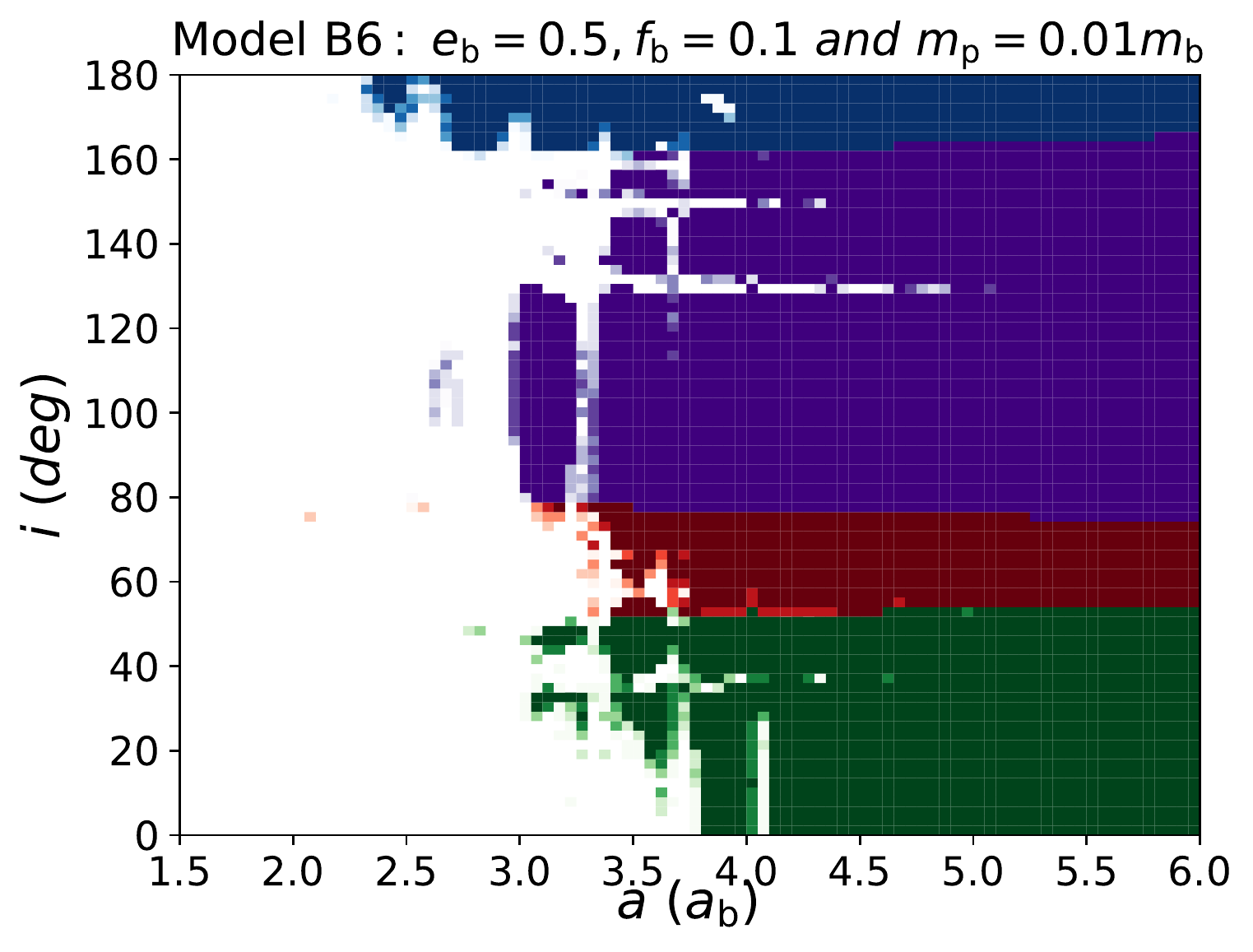}
    \includegraphics[width=5.8cm]{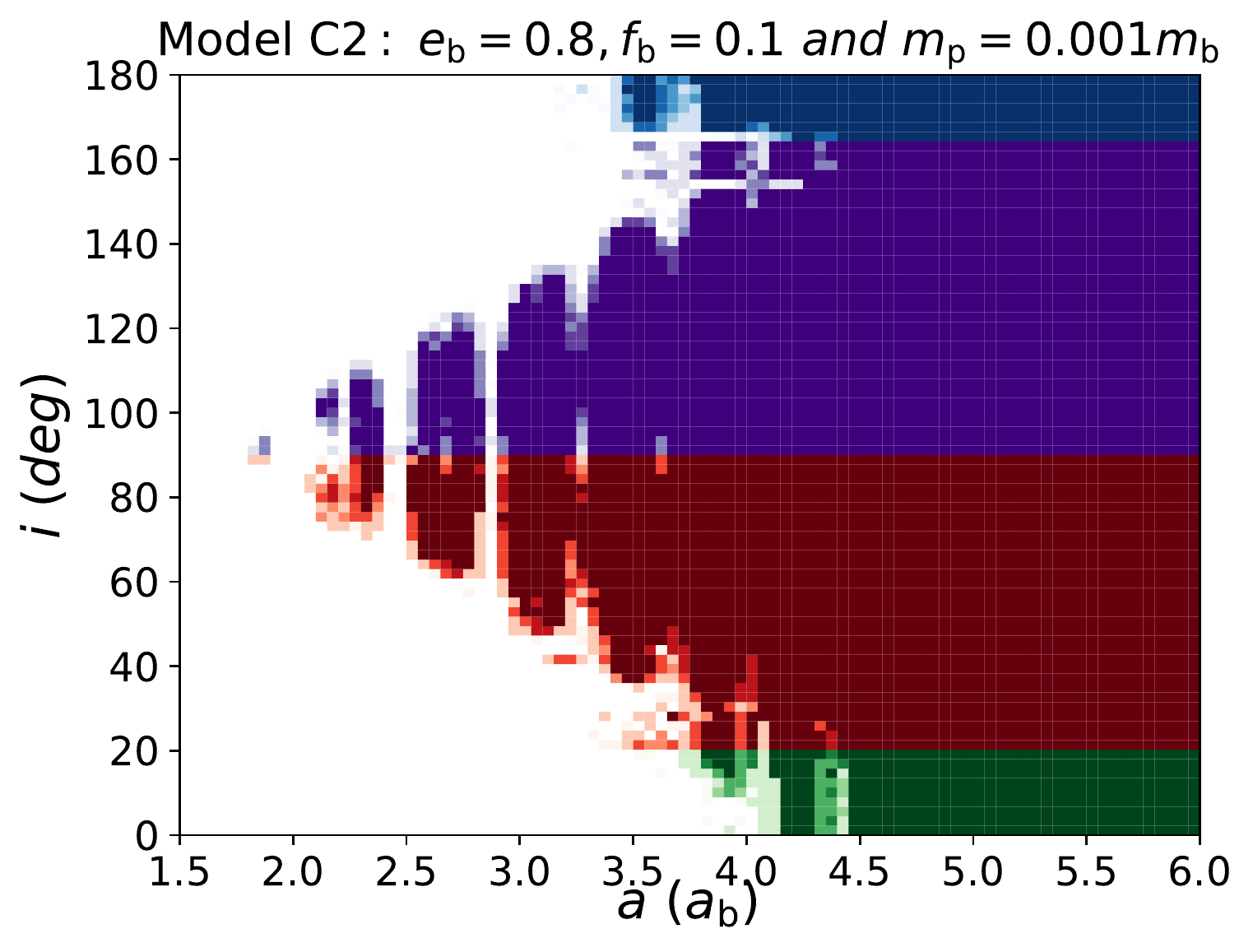}
    \includegraphics[width=5.8cm]{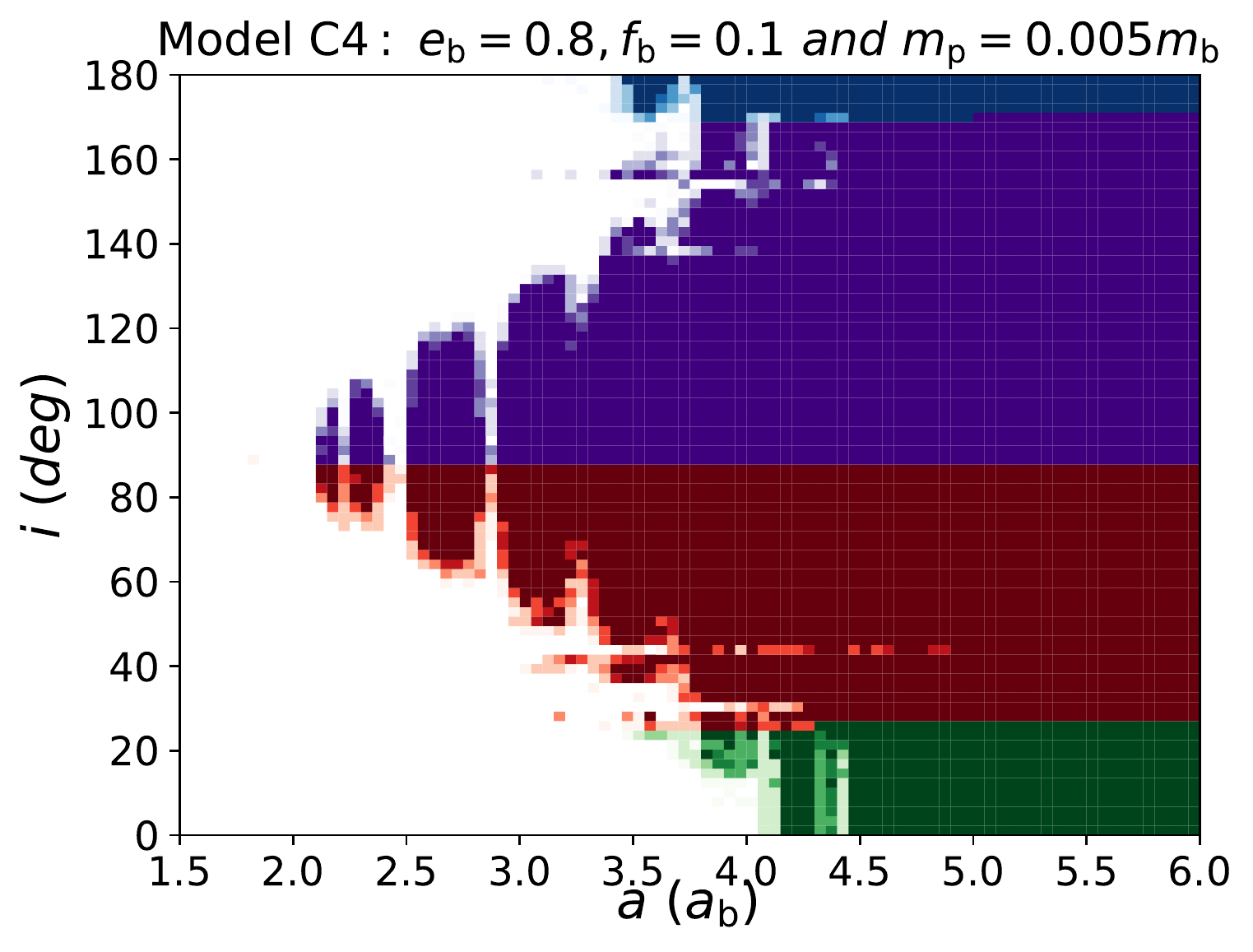}
    \includegraphics[width=5.8cm]{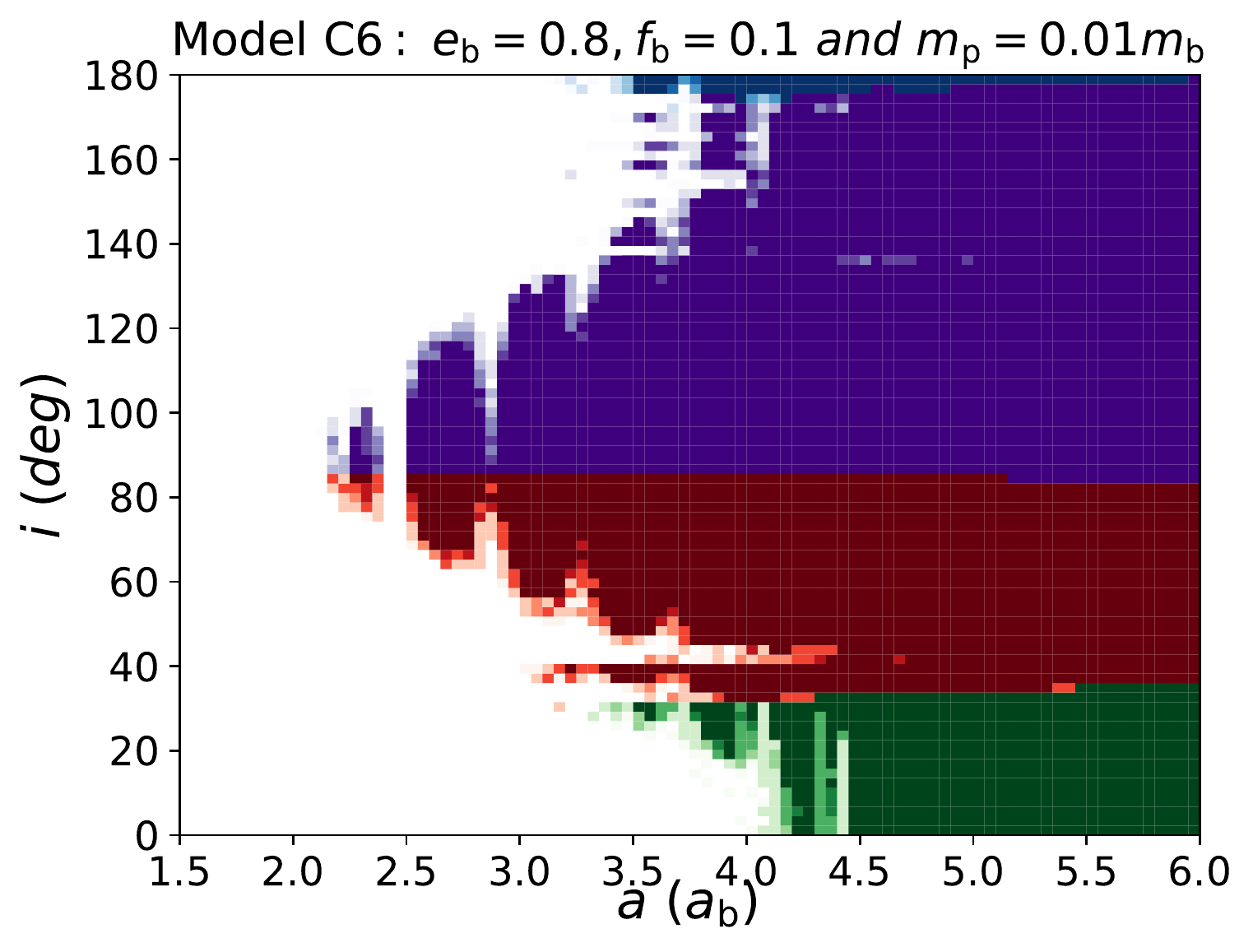}
    \caption{Same as Fig~\ref{fig:map1} except the binary mass fraction is $f_{\rm b}$=0.1.}
    \label{fig:map2}
\end{figure*}

 We first consider the stability maps for equal mass binaries and then for a binary mass fraction of $f_{\rm b}=0.1$.

\subsubsection{Equal mass binary}

Fig.~\ref{fig:map1} shows stability maps for equal mass binary models with initial binary eccentricity $e_{\rm b}=0.2$ (models A1, A3, and A5, first row), models with binary eccentricity $e_{\rm b}=0.5$ (models B1, B3, and B5, second row), and models with binary eccentricity $e_{\rm b}=0.8$ (models C1, C3, and C5, third row). In each panel, we vary the initial separation, $a$, and the initial inclination, $i$. For each combination of these two parameters considered, we vary the true anomaly $\nu$ from 0 to  $300^{\circ}$ with an interval of $\Delta= 60^{\circ}$. Thus, each pixel in Fig.~\ref{fig:map1} represents six simulations in total. 

The colour of each pixel represents the type of planet orbit. We follow the colour representation in \citet{Chen20192}, except we use purple  instead of cyan. Thus, the prograde circulating orbits are green and the retrograde circulating orbits are blue. The librating orbits with initial inclination $i<i_{\rm s}$ are red and those with initial inclination $i>i_{\rm s}$ are purple. Notice that one pixel can only display one colour for the six simulations in each pixel.
 In some cases, there may be two types of orbits in the same pixel. In such cases, the colour refers to the type of orbit that has the largest number of the six stable orbits.
 The darker the colour, the larger the number of stable  orbits. 
White pixels indicate unstable orbits.  A lighter nonwhite pixel indicates that some contributing
orbits are unstable.

As seen in Fig.~\ref{fig:map1}, the prograde circulating (green) orbits are the generally least stable orbits for the equal mass binary, meaning that the minimum initial semi--major axis $a$ for stable orbits is generally larger than for the other orbit types. The critical inclination below which the orbit is prograde circulating (green) (i.e., the inclination at red--green boundary) is higher for smaller binary eccentricity and higher planet mass. The critical inclination does not vary much with planet separation for the small separations considered, as expected in the analytic model of \cite{MartinandLubow2019}. For these green orbits, the minimum stable initial planet semi--major axis does not change much with planet inclination or planet mass, but is more strongly affected by the binary eccentricity. The higher the binary eccentricity the more unstable the prograde circulating (green) orbits are.

In Fig.~\ref{fig:map1}, the retrograde circulating (blue) orbits are the most stable orbits for low binary eccentricity $e\lesssim 0.5$.  The critical inclination above which the orbit is retrograde circulating (blue) (i.e., inclination at the purple--blue boundary) is lower for smaller binary eccentricity and lower planet mass. These circulating retrograde orbits are generally stable closer to the binary than the circulating prograde orbits. But there is significant dependence of stability on the initial planet orbit inclination with orbits starting closer $i=180^{\circ}$ being more stable. 
For the equal mass binary case shown here, there is not much difference in stability of the retrograde circulating orbits across different values of the planet mass (i.e., across a row in Fig.~\ref{fig:map1}). 

The librating orbits are the most stable orbits for high binary eccentricity. The stationary inclination decreases with increasing planet mass. That is, the inclination $i_{\rm s}$ at the red--purple boundary decreases with increasing planet mass, as expected in the analytic model of \cite{MartinandLubow2019}. The range of inclinations for which there are librating orbits with initial $i<i_{\rm s}$ (red orbits) decreases with planet mass while the range of inclinations for which librating orbits with initial $i>i_{\rm s}$ (purple orbits) increases with planet mass. The most stable orbits at high eccentricity occur near the stationary inclination. Generally, the larger the difference in the initial planet inclination from the stationary inclination $|i-i_{\rm s}|$, the more unstable the librating planet orbits.

Comparing with Figure 14 in \citet{Doolin2011}, we find that our results are in general agreement with their test particle simulations, since the mass of the planet has not affected our results much for an equal mass binary. However, the stationary inclination, $i_{\rm s}$, is always 90$^\circ$ in their simulations because the test particle does not have angular momentum to exchange with the binary system. In addition, we have considered higher binary eccentricity than their simulations that cover up to $e_{\rm b}=0.6$. Consequently, we see more clearly that the near polar librating orbits are  the most stable at high binary eccentricity
for an equal mass binary.

\begin{figure*}
    \centering
    \includegraphics[width=5.8cm]{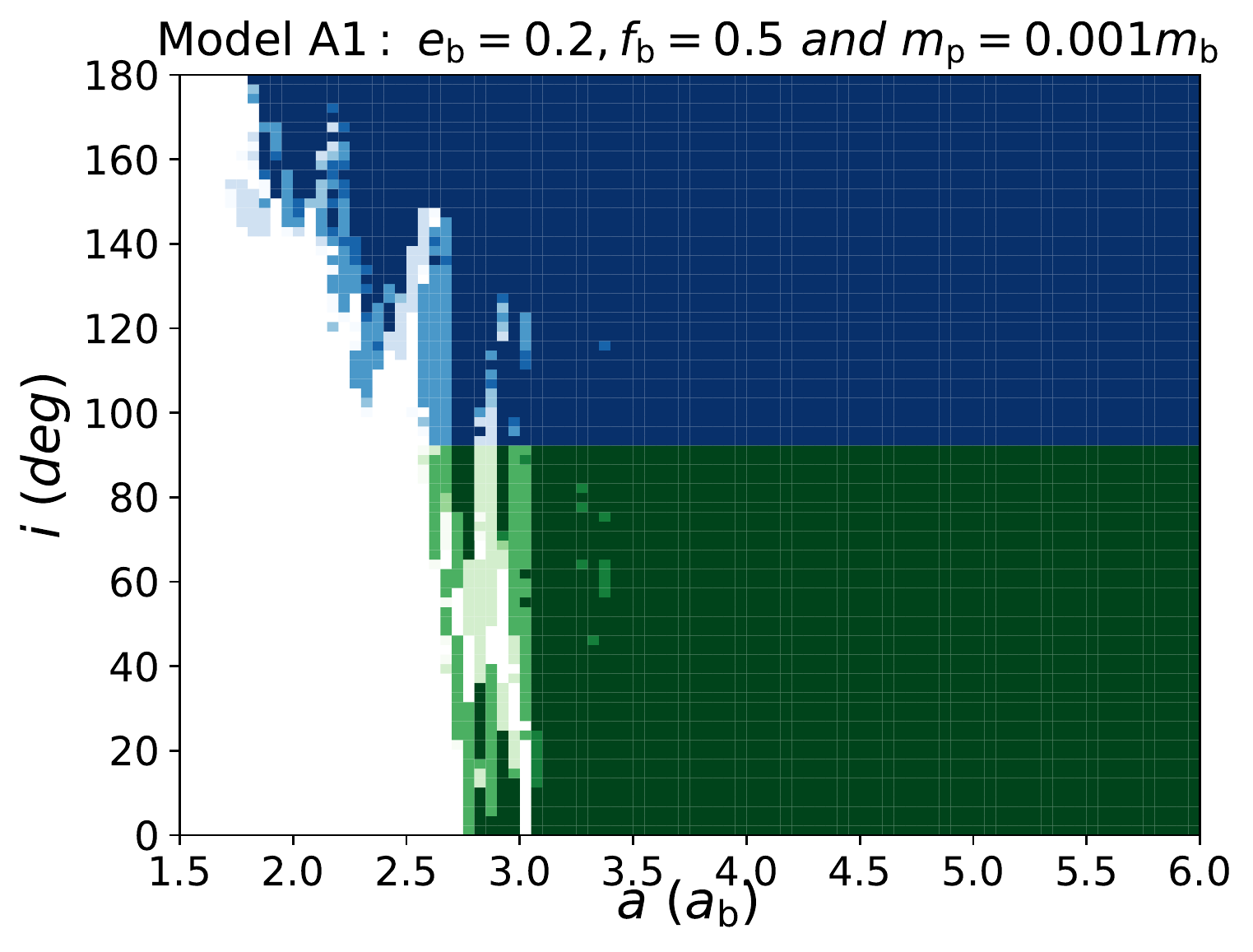}
    \includegraphics[width=5.8cm]{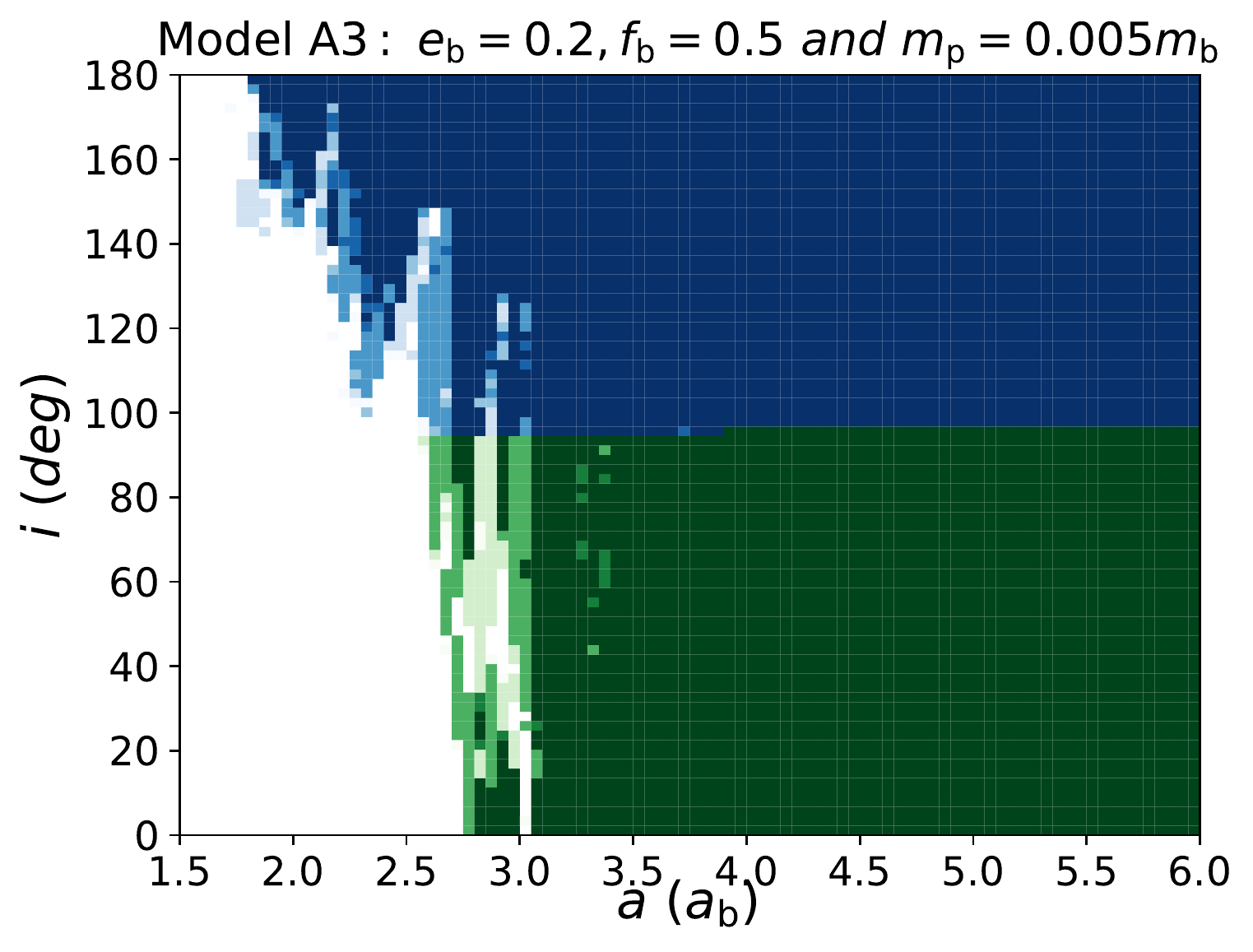}
    \includegraphics[width=5.8cm]{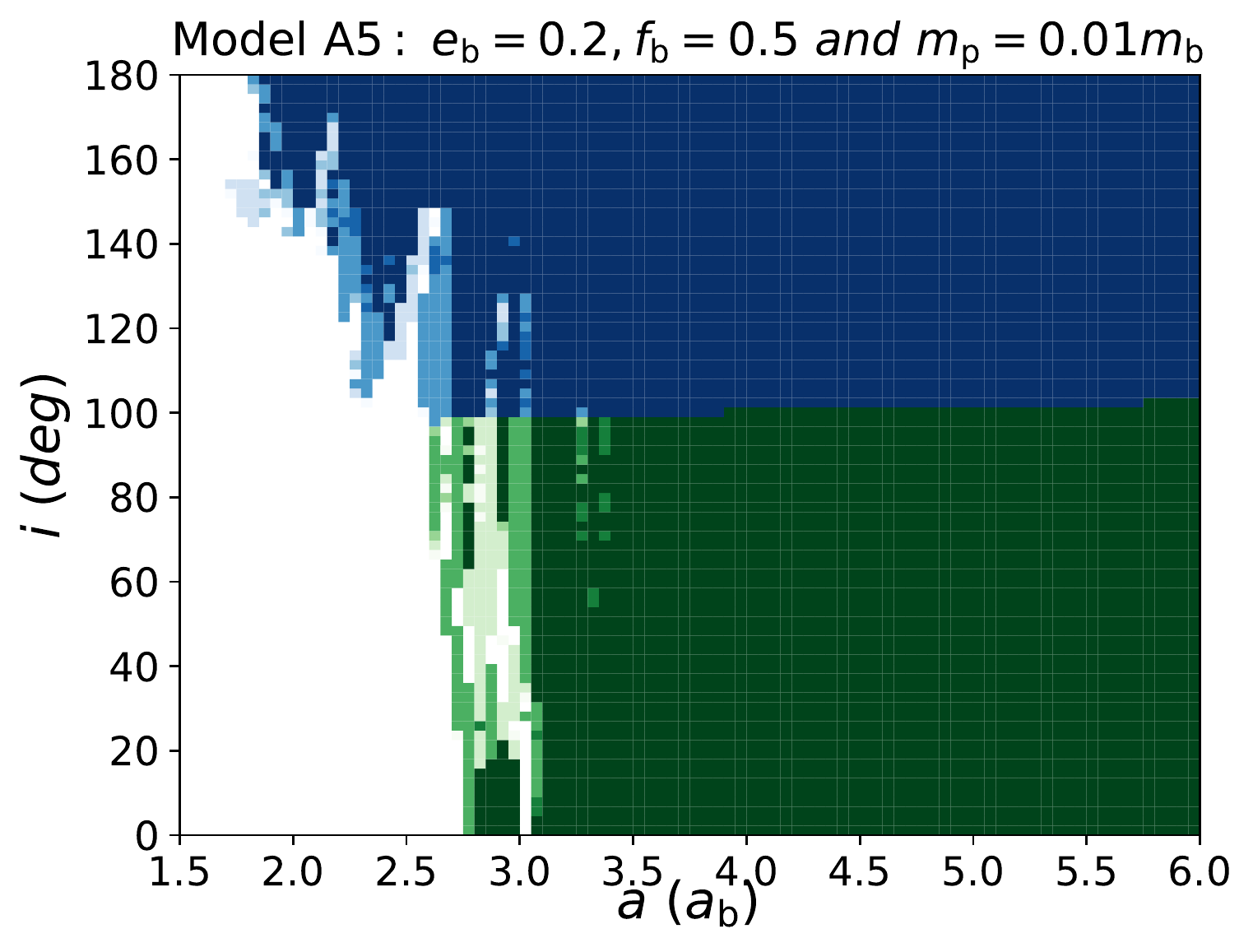}
    \includegraphics[width=5.8cm]{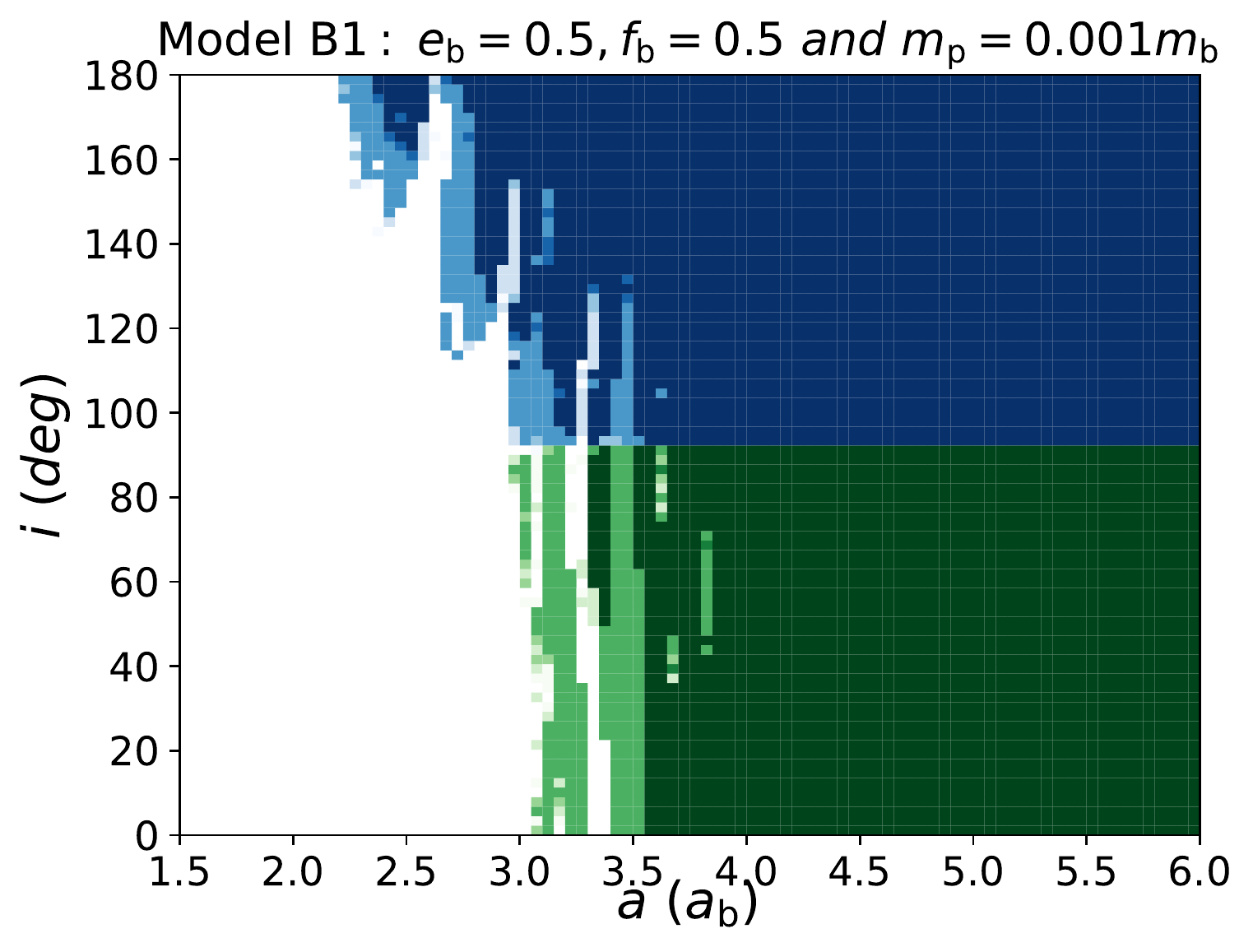}
    \includegraphics[width=5.8cm]{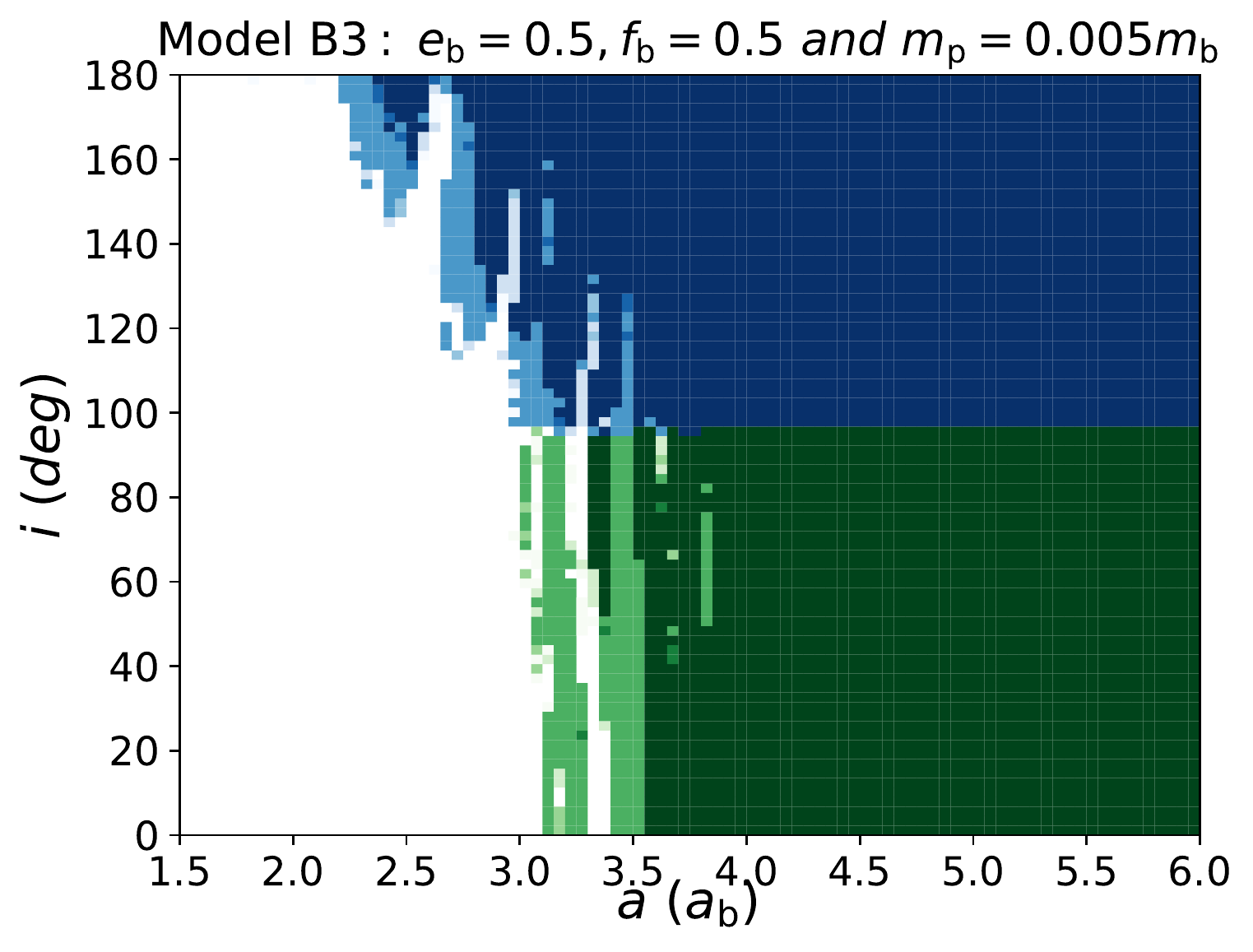}
    \includegraphics[width=5.8cm]{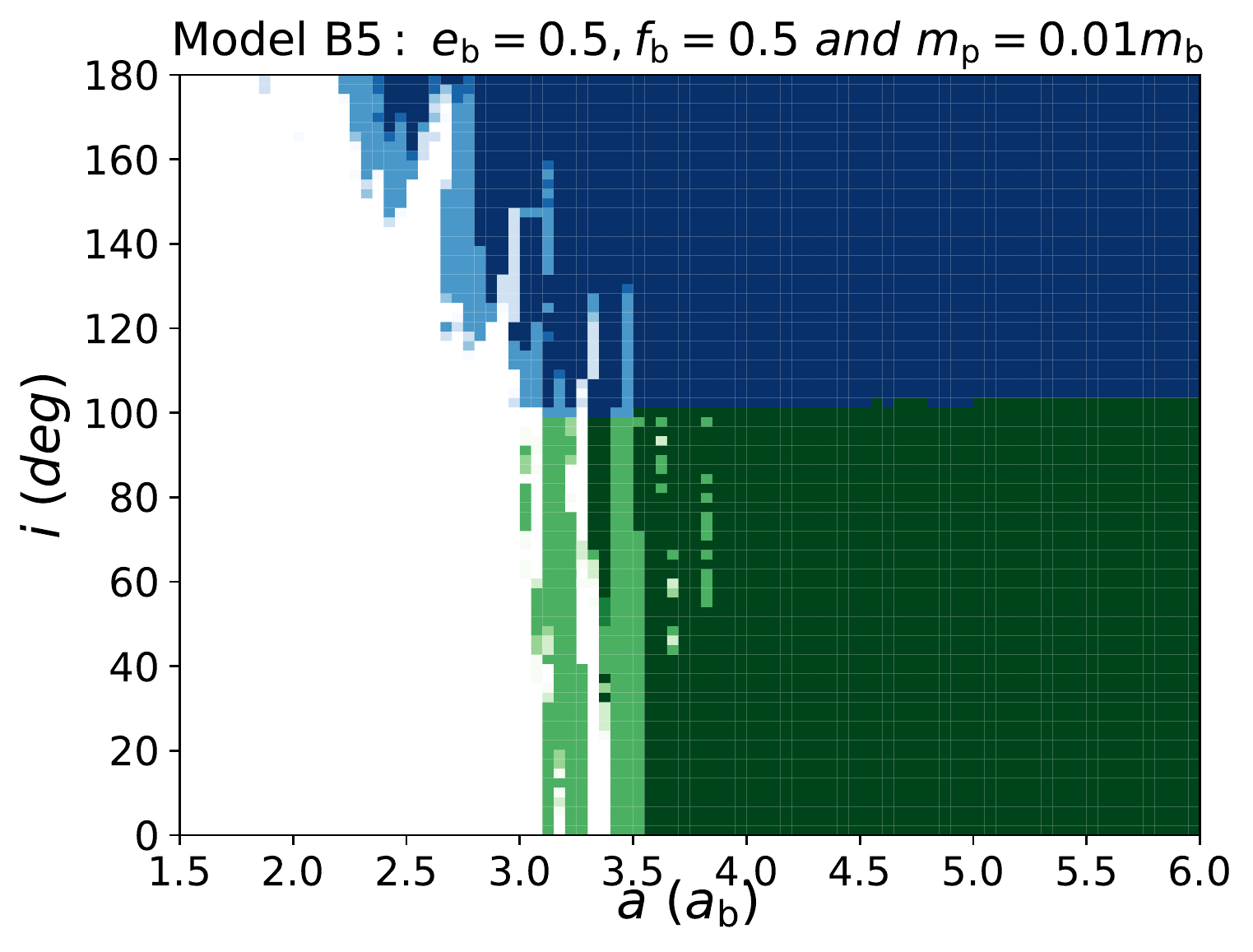}
    \includegraphics[width=5.8cm]{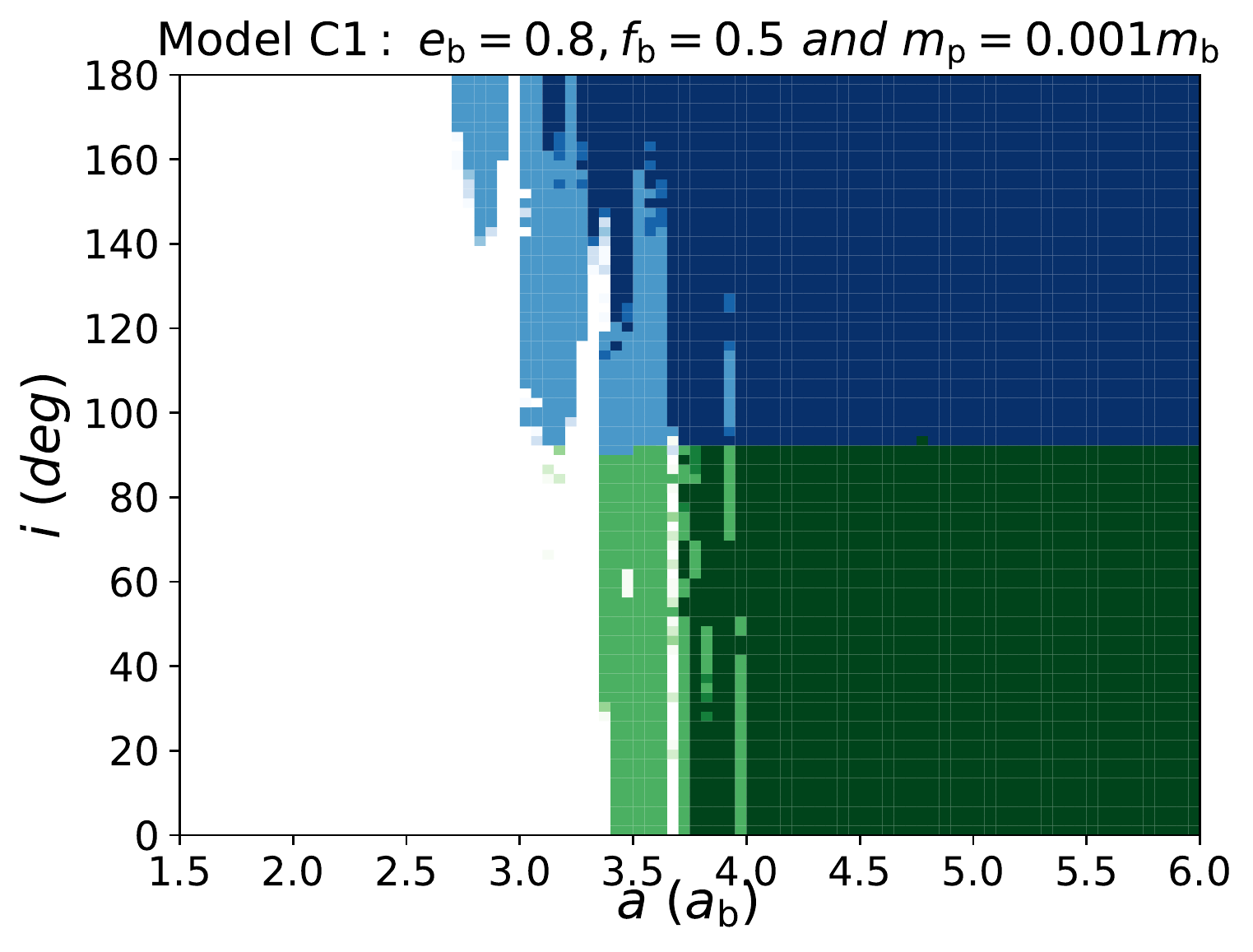}
    \includegraphics[width=5.8cm]{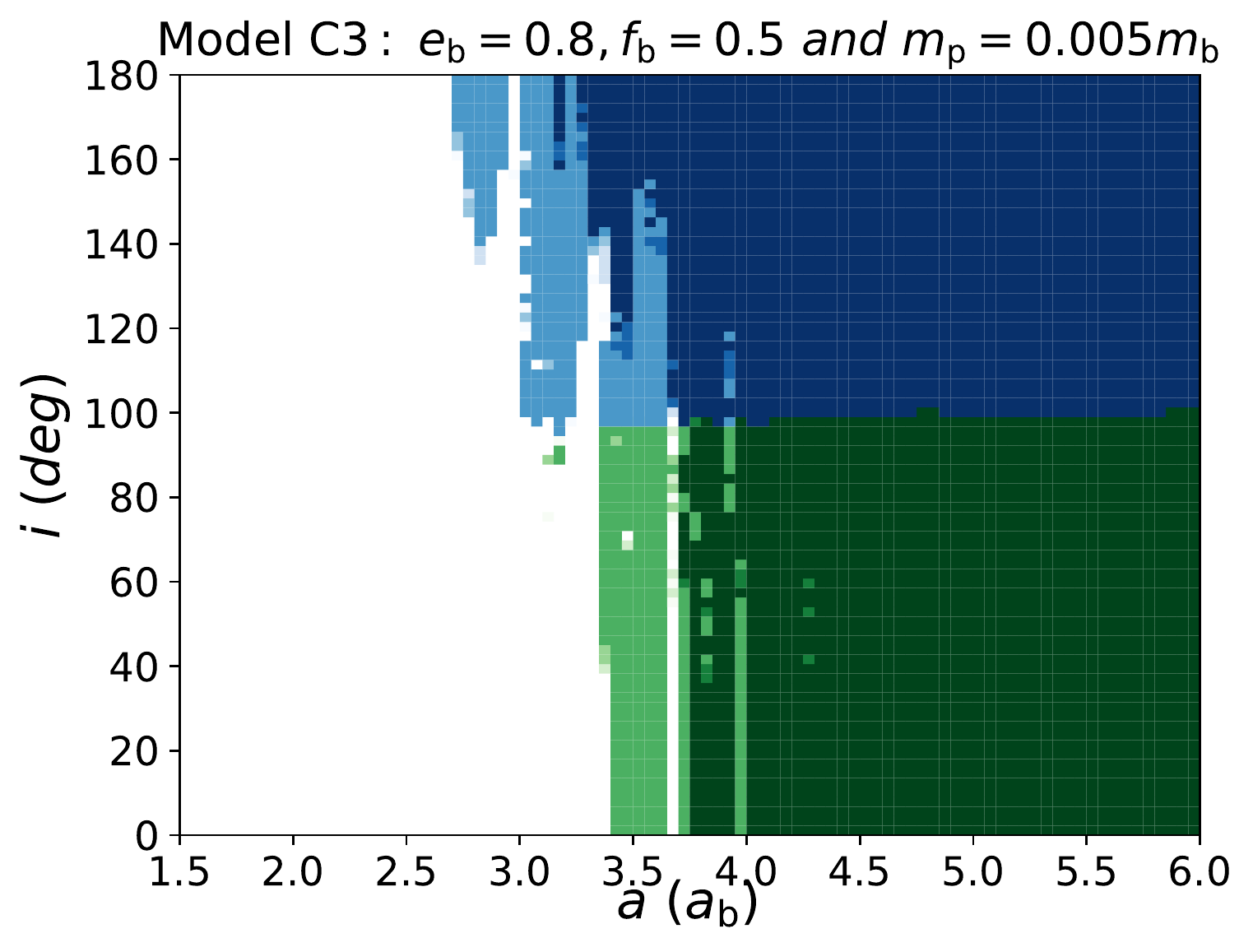}
    \includegraphics[width=5.8cm]{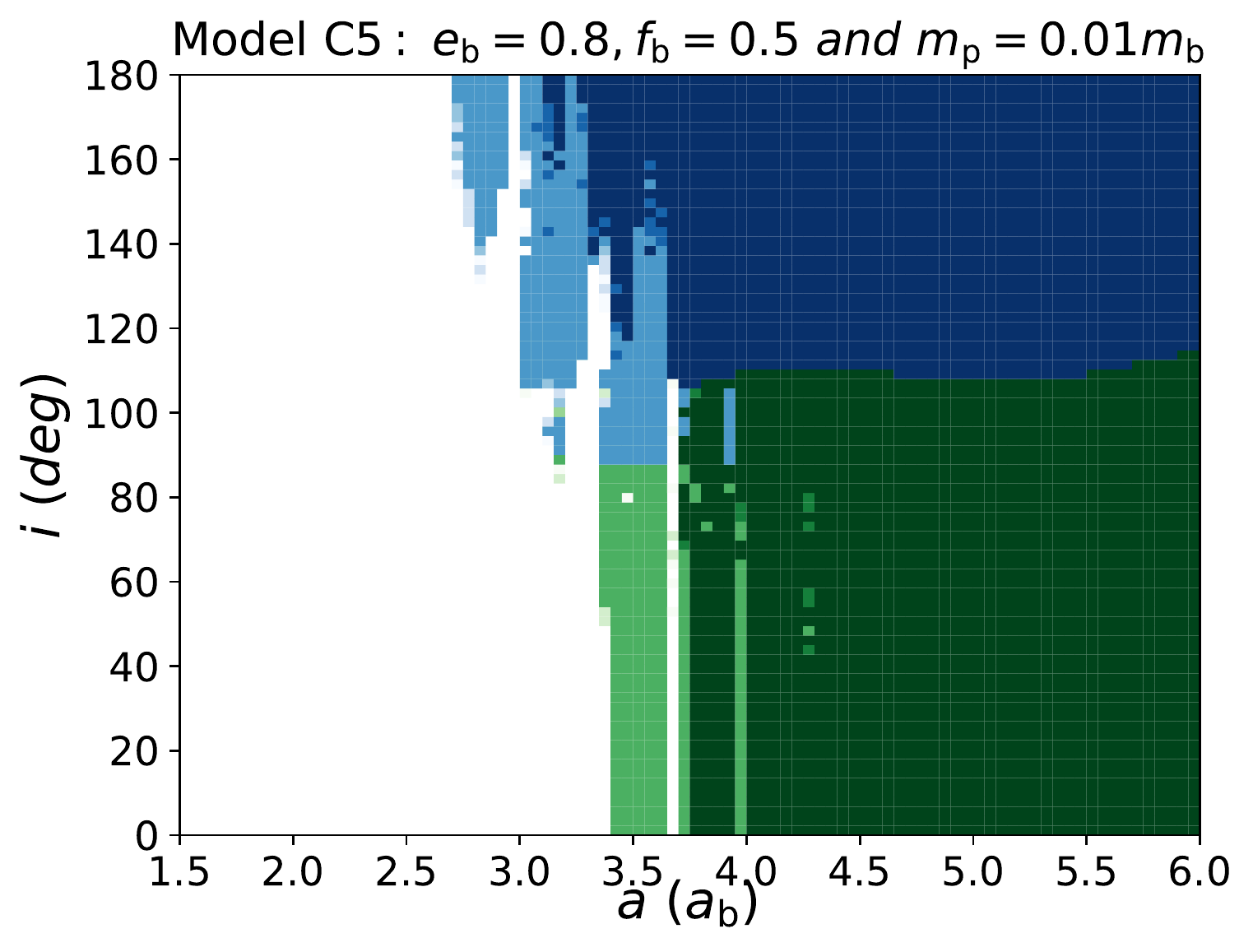}
    \caption{Same as Fig.~\ref{fig:map1} except initial $\phi=0^\circ$.}
    \label{fig:phi=0fb=5}
\end{figure*}

\begin{figure*}
    \centering
    \includegraphics[width=5.8cm]{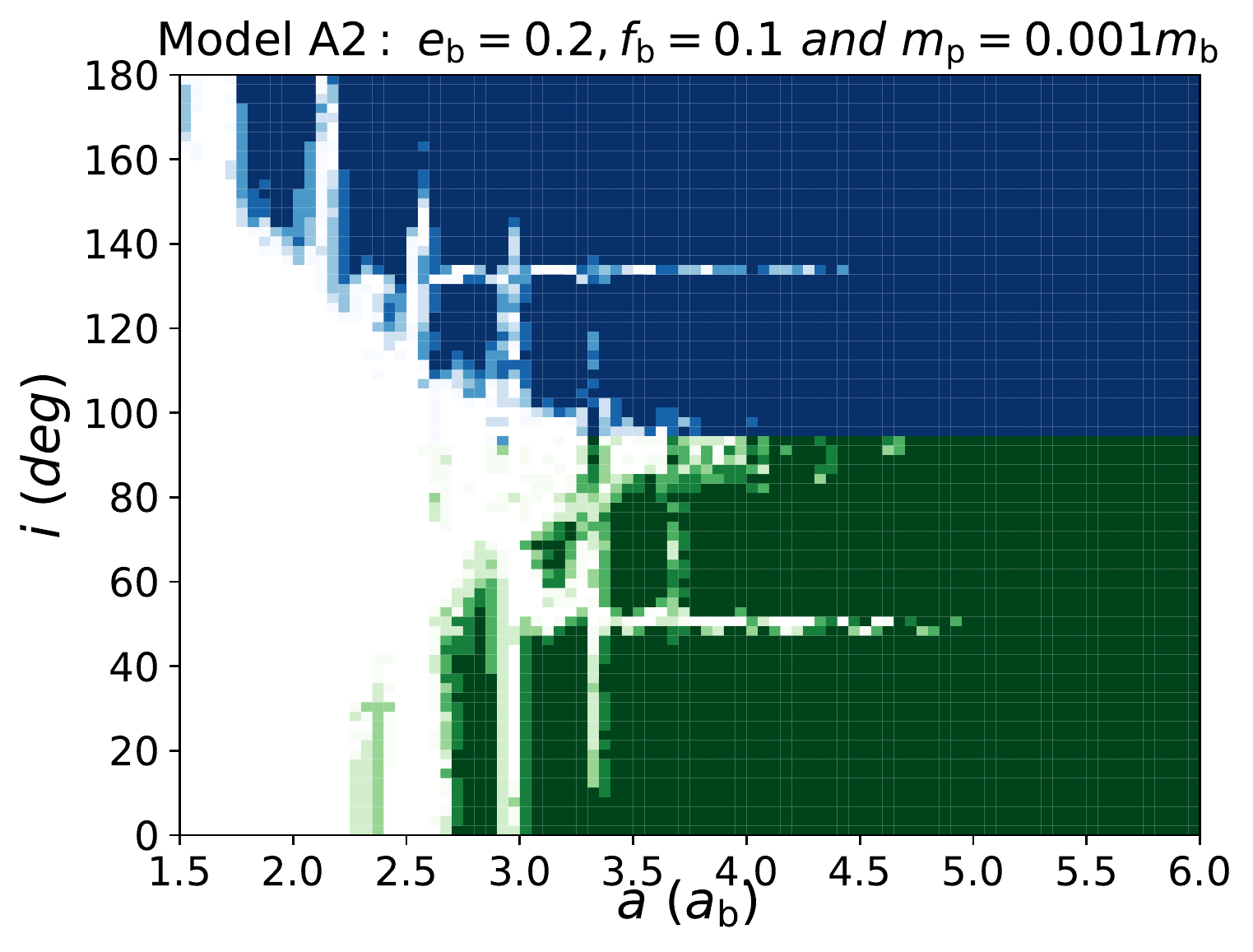}
    \includegraphics[width=5.8cm]{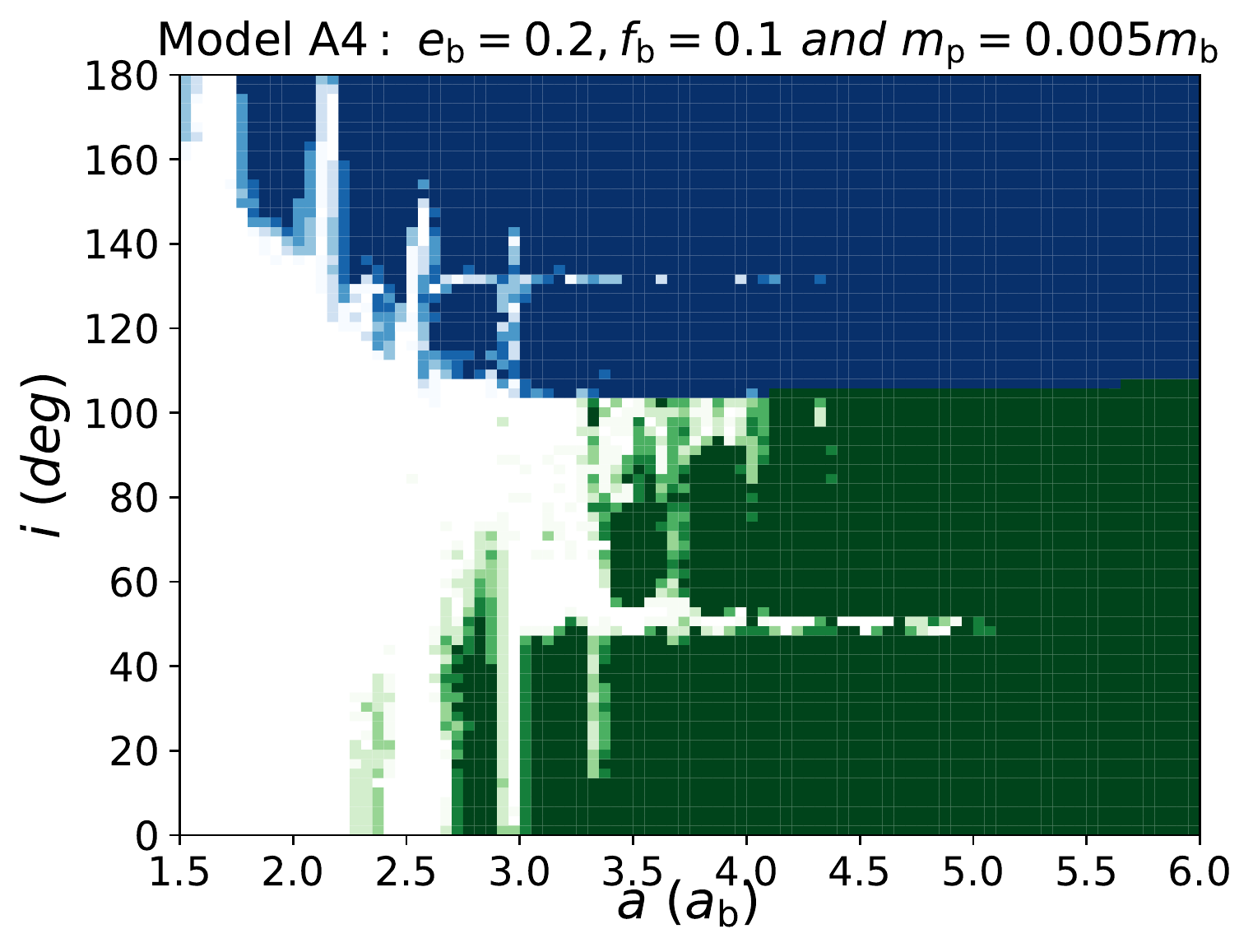}
    \includegraphics[width=5.8cm]{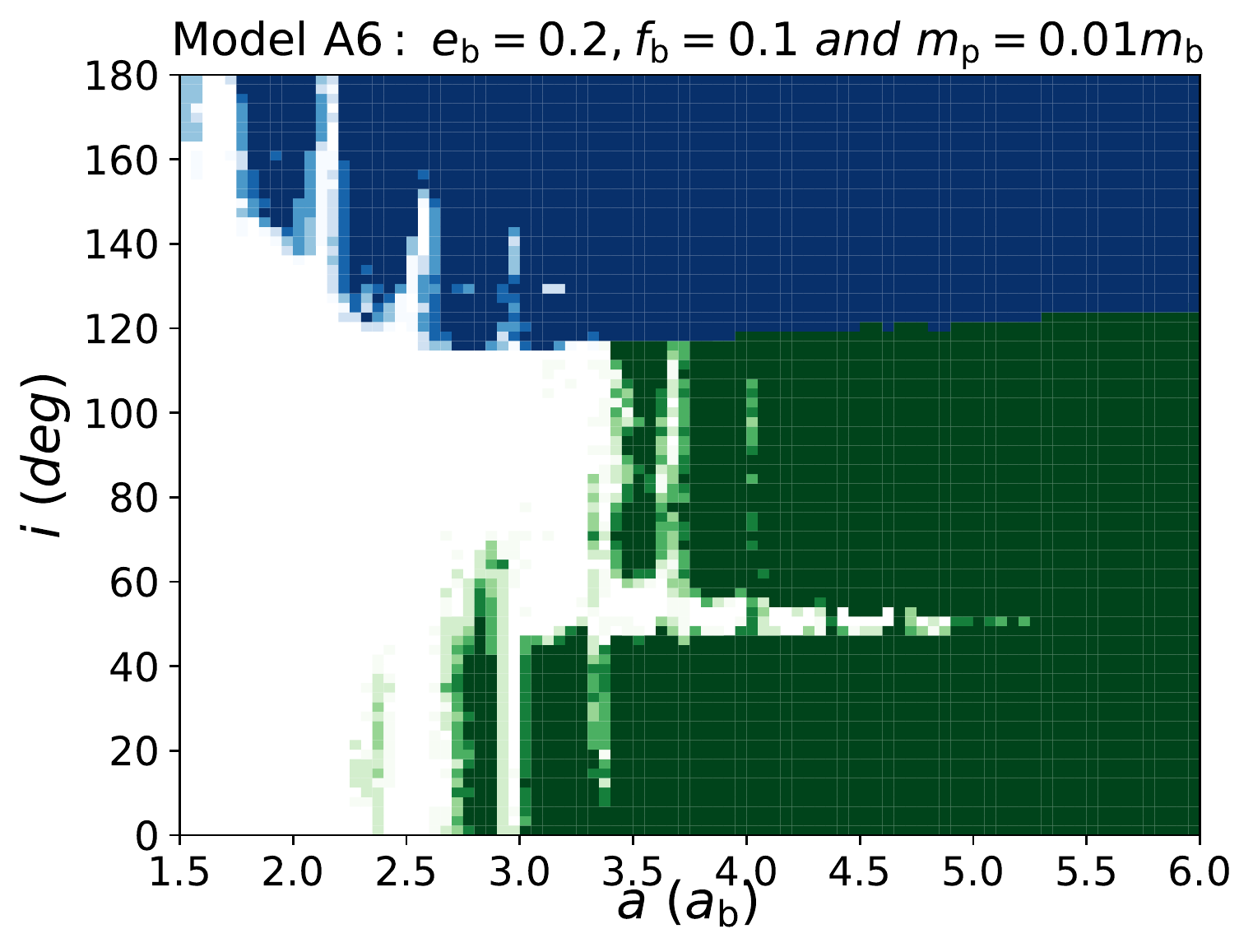}
    \includegraphics[width=5.8cm]{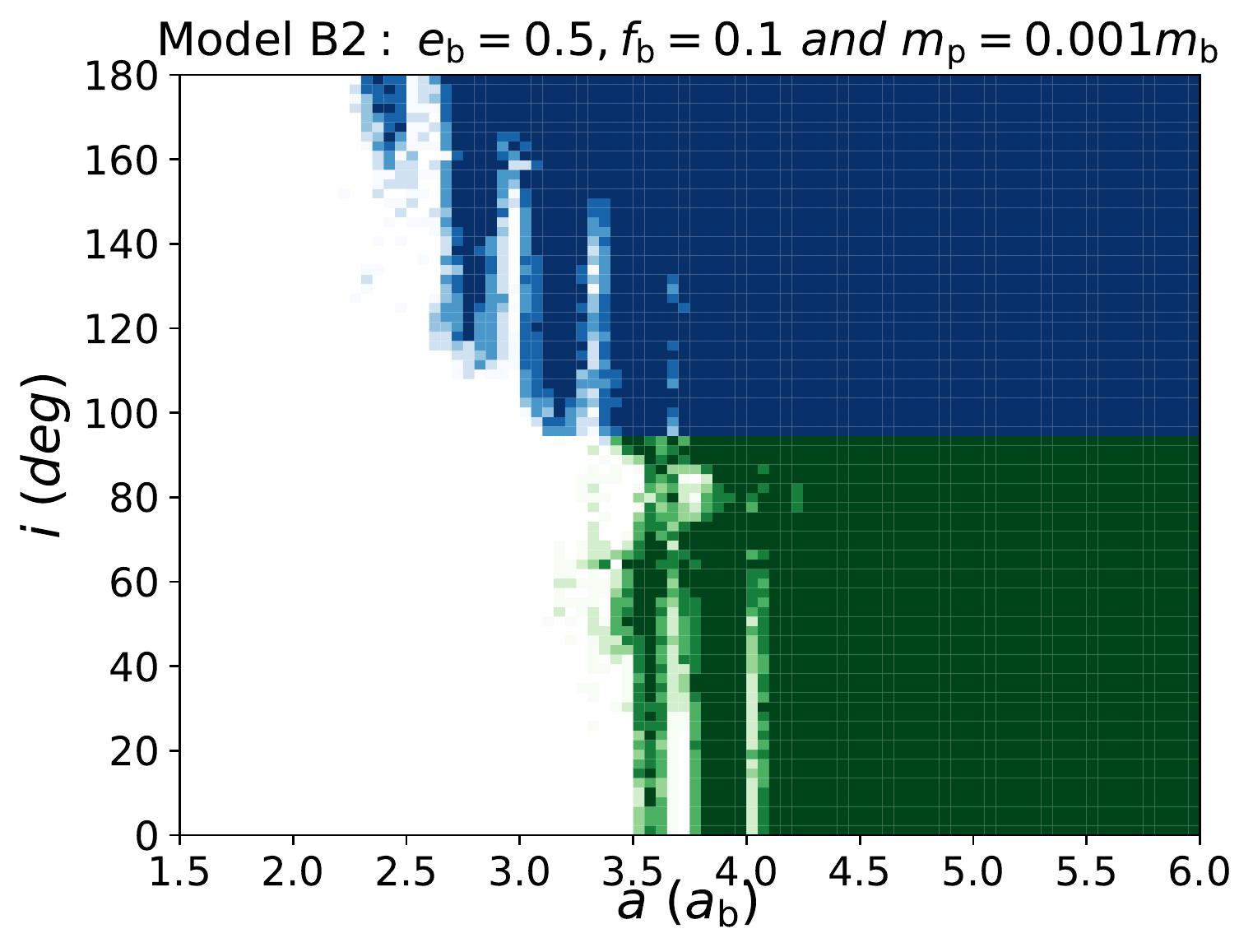}
    \includegraphics[width=5.8cm]{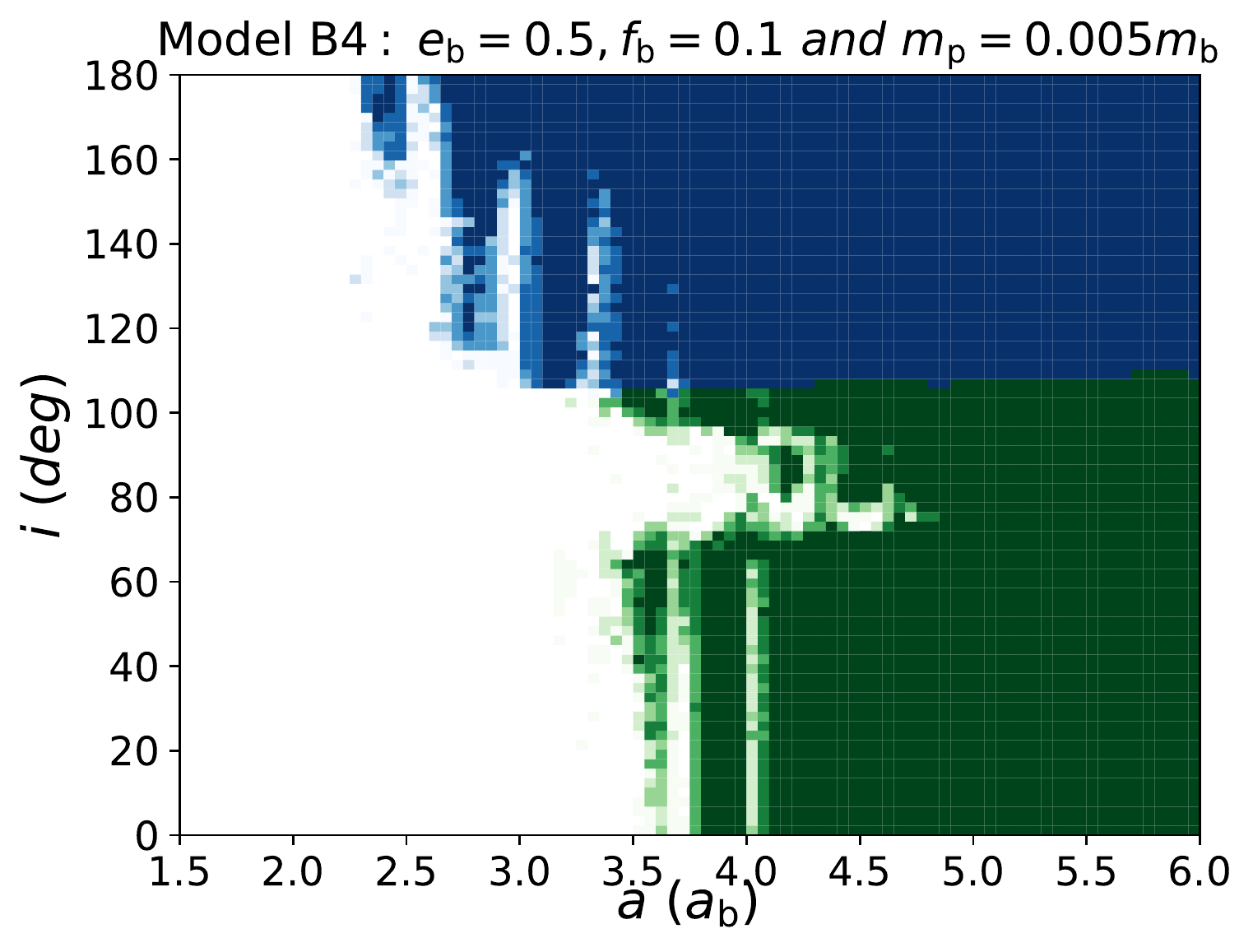}
    \includegraphics[width=5.8cm]{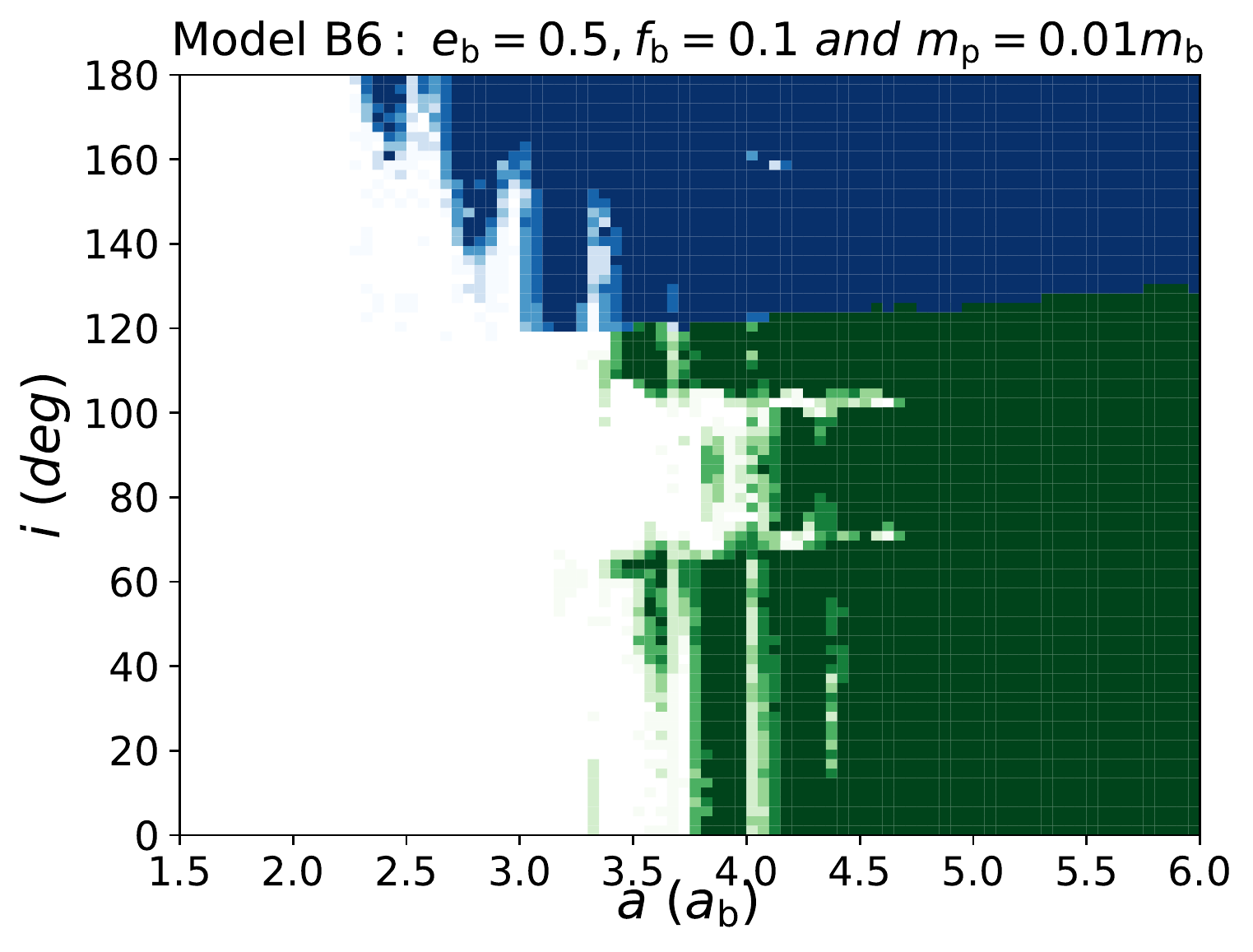}
    \includegraphics[width=5.8cm]{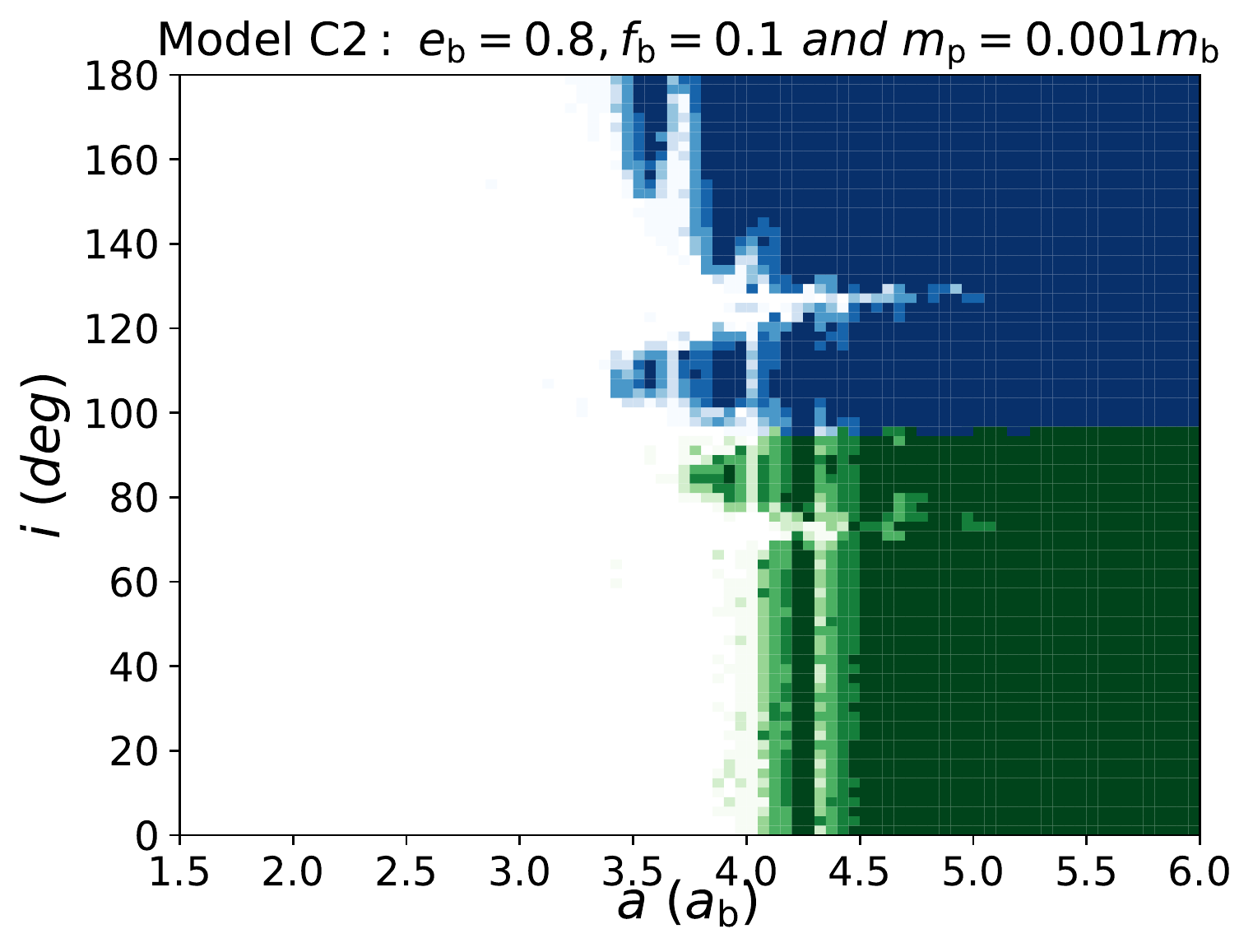}
    \includegraphics[width=5.8cm]{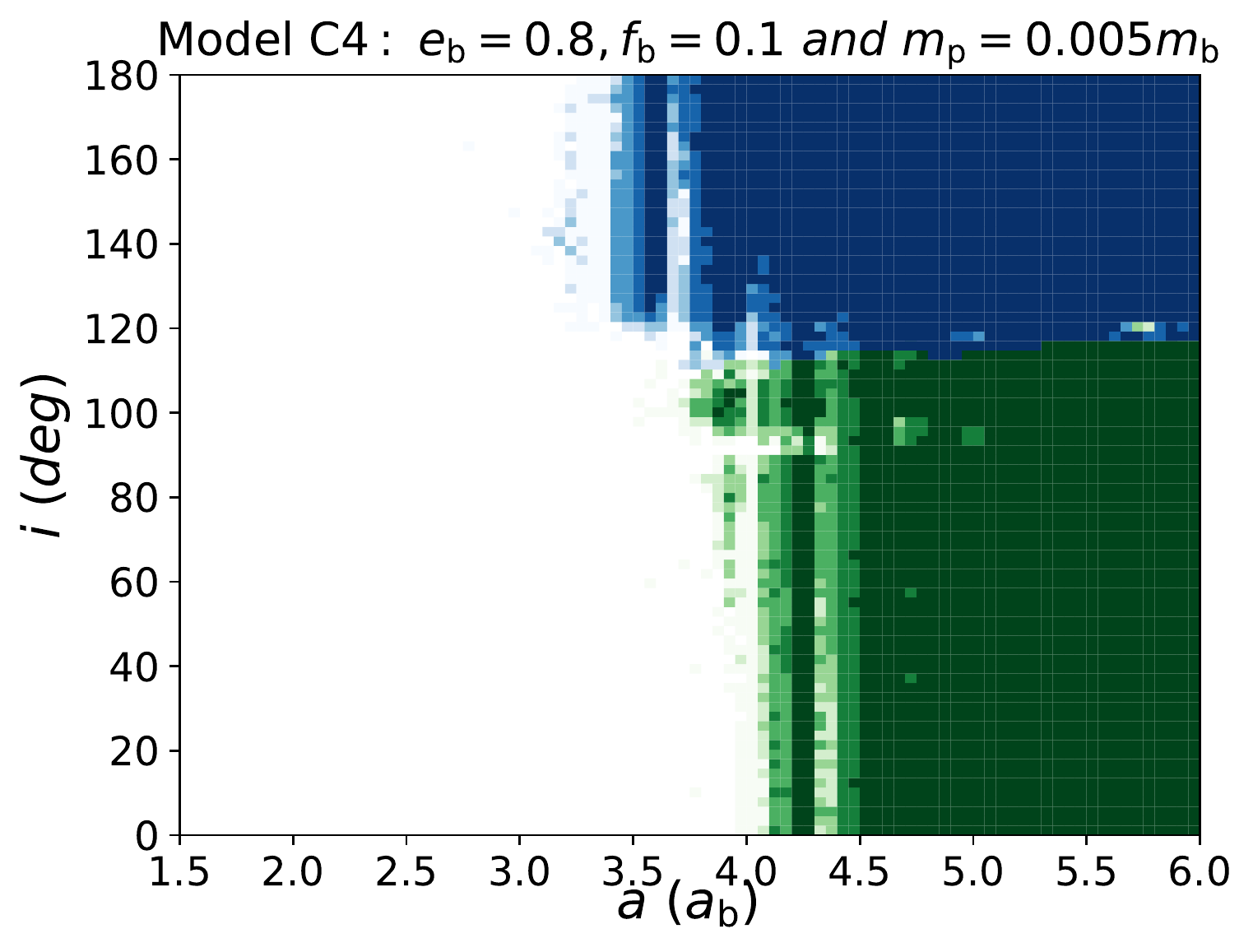}
    \includegraphics[width=5.8cm]{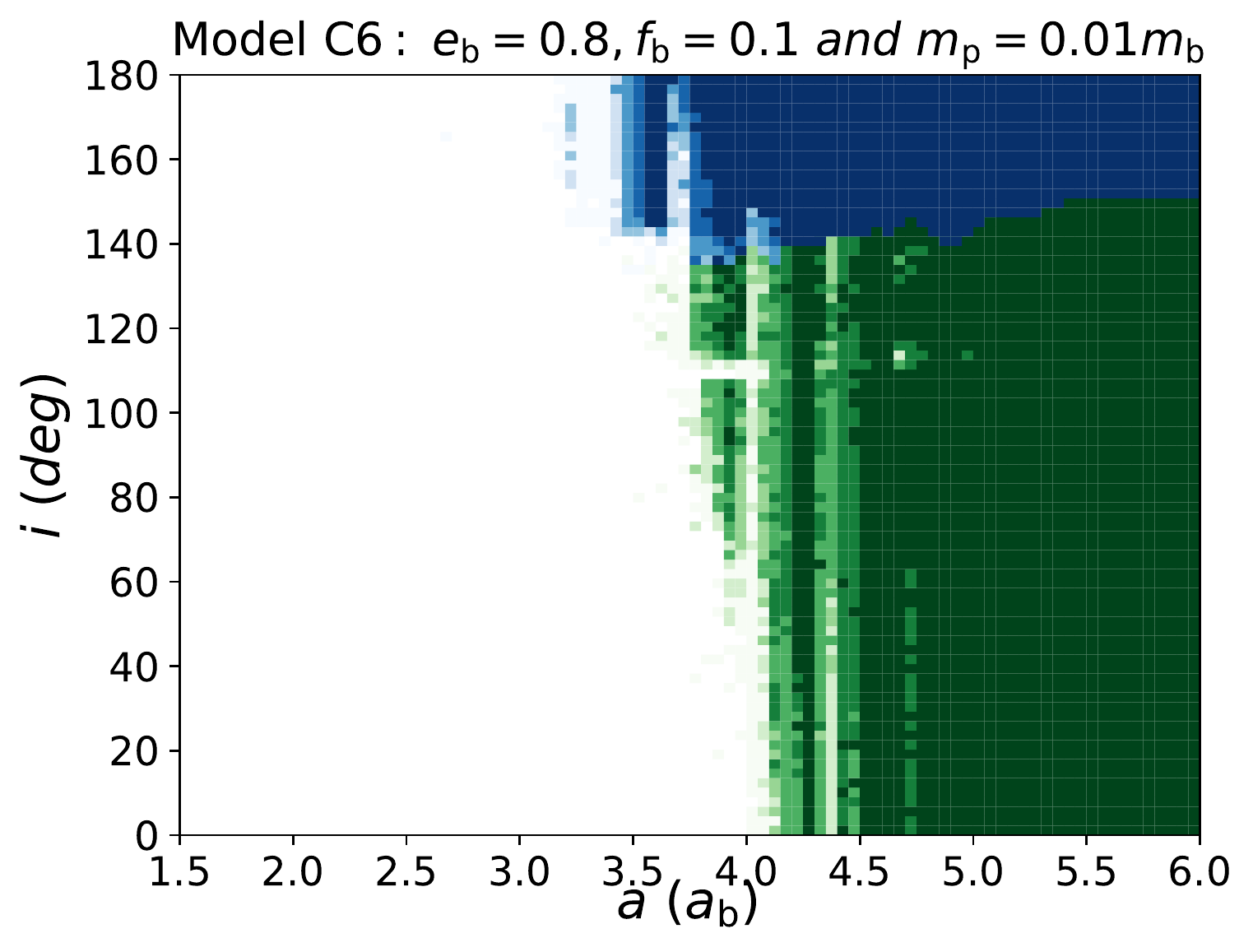}
    \caption{Same as Fig.~\ref{fig:map2} except initial $\phi=0^\circ$.}
    \label{fig:phi=0}
\end{figure*}

\subsubsection{Binary with mass fraction $f_{\rm b}=0.1$}

Fig.~\ref{fig:map2} shows the stability maps for binary mass fraction $f_{\rm b}=0.1$ for binary eccentricity $e_{\rm b}=0.2$ (models A2, A4, and A6, first row),  binary eccentricity $e_{\rm b}=0.5$ (models B2, B4, and B6, second row) and  binary eccentricity $e_{\rm b}=0.8$ (models C2, C4, and C6, third row). The range of angles for which the planet is librating is larger compared with the equal mass binary case, however there is much more instability for low binary mass fraction. The effect of increasing the mass of the planet is much more significant for $f_{\rm b}=0.1$ than it is for an equal mass binary,  as a consequence of the planet mass being closer to the binary secondary mass.

The prograde circulating (green) orbits are the least stable orbits for high binary eccentricity, $e_{\rm b} \geq 0.5$, similar to the equal mass binary case. The innermost stable prograde orbits have $a \simeq 2.3\,a_{\rm b}$ for the simulation with $e_{\rm b} =0.2$.  This radius does not change much with the mass of planet for low binary eccentricity. However, the innermost stable orbits are much farther out for high binary eccentricity, the separation is $a \simeq 3.5 \,a_{\rm b}$ for binary eccentricity $e_{\rm b}=0.8$. The critical inclination  below which the orbit is prograde and circulating (i.e., inclination at the green--red boundary) increases by about 3$^{\circ}$ as the planet mass increases for $e_{\rm b}$ = 0.2, while for $e_{\rm b}$ = 0.8, it increases by about 15$^\circ$ with planet mass increasing from $0.001\,m_{\rm b}$ up to $0.01\,m_{\rm p}$. These results are consistent with analytic solutions of the green-red lines in the bottom panels of Fig.~8 in \citet{Chen20192} that show that a low mass planet has higher $i_{\rm crit}$ for small $e_{\rm b}$  and lower $i_{\rm crit}$ for large $e_{\rm b}$. The boundary between  prograde circulating and the librating orbits is unstable even if the planet is at large separation, $a > 5 \,a_{\rm b}$, in the high mass model ($m_{\rm p}$ = 0.01 $m_{\rm b}$) with low binary eccentricity $e_{\rm b} = 0.2$ (Model A6). However, with increasing binary eccentricity, Model B6 and Model C6 show that the boundary becomes more stable.

For low binary eccentricity $e_{\rm b} \leq 0.5$, the retrograde circulating (blue) orbits are again the most stable orbits. The innermost stable retrograde region is at separation $a \simeq 1.5 \,a_{\rm b}$ for $e_{\rm b}=0.2$ and  $a \simeq 3.2 \,a_{\rm b}$ for $e_{\rm b} =0.8$. The critical inclination above which the orbit is retrograde circulating (i.e., inclination at the purple-blue boundary) is lower for smaller $e_{\rm b}$ and lower $m_{\rm p}$. It increases by about 20$^\circ$ with increasing $m_{\rm p}$ for all the binary eccentricities considered here. For the high $e_{\rm b}$ model, the critical inclination of the retrograde circulating orbits increases to very close $180^{\circ}$ for $m_{\rm p} = 0.01\, m_{\rm b}$. This is consistent with blue dotted lines in the two bottom panels of Fig. 8 in \citet{Chen20192} that show the critical angles of the low mass planet and the high mass planet increase by about 20$^\circ$ as the binary eccentricity increases from $e_{\rm b}$ = 0.2 to 0.8 in the binary system with $f_{\rm b}$ = 0.1.

As in the equal mass binary case, the librating orbits are the most stable orbits for high binary eccentricity. But in this  case of an unequal mass binary, the difference in stability between librating
and circulating orbits is greater than in the equal mass binary case, especially at high binary eccentricity.
The innermost stable libration region is at $a \simeq 2.5 \,a_{\rm b}$ for $e_{\rm b} =0.2$ and  $a \simeq 2.3 \,a_{\rm b}$ for $e_{\rm b} =0.8$ for the models with $m_{\rm p}$ = 0.01 $m_{\rm b}$.  The stationary inclination $i_{\rm s}$ decreases with increasing $m_{\rm p}$. Thus, the inclination of the transition from red to purple orbits $i_{\rm s}$  decreases with increasing planet mass. For the small $e_{\rm b}$ models  ($e_{\rm b}$ = 0.2), the purple libration region expands with increasing planet mass $m_{\rm p}$, while the red libration region becomes very small. Thus, there are no red librating orbits in the $i\cos \phi$--$i \sin \phi$ phase plot of Model D1 in Fig. 3 in \citet{Chen20192}. 

For the test particle orbits considered in \cite{Doolin2011}, the librating orbits in the stablility maps appear close to symmetric about the stationary inclination $i=i_{\rm s}=90^\circ$. However, the symmetry is broken for a planet with mass. The asymmetry is larger for lower $e_{\rm b}$ and higher $m_{\rm p}$. For the low $m_{\rm p}$ models, even for a planet at large radius ($a > 5 a_{\rm b}$), the purple librating orbits with $i$ = $100^{\circ} < i < 120^{\circ}$ are unstable while the red libration region is stable. This is consistent with the $i\cos \phi$--$i \sin \phi$ phase plot of Model B1 in the Fig. 1 in \citet{Chen20192} where there were no stable orbits for initial $i$ in this region. On the other hand, for the high $m_{\rm p}$ model at small binary eccentricity $e_{\rm b}=0.2$, the orbit of the planet at large radius ($a > 5 a_{\rm b}$) with $i = 60^{\circ} < i < 80^{\circ}$ is unstable. This is consistent with the $i\cos \phi$--$i \sin \phi$ phase plot of Model D1 in the Fig. 3 in \citet{Chen20192} where there were no red stable librating orbits for initial $i$ in this range.

Overall, for initial $\phi=90^\circ$, at low binary eccentricity, the polar orbits are generally more unstable for the low mass fraction binary compared to the case of an equal mass binary. The retrograde circulating orbits are the most stable orbits for low $e_{\rm b}$ and the innermost region of the stable retrograde orbit can extend down to $a = 1.5 \,a_{\rm b}$. The mass of the planet does not affect the stability map much for equal mass binary, but it has a significant impact for the unequal mass binary, particularly for low binary eccentricity. For high binary eccentricity, the librating orbits are generally the most stable, particularly for inclinations close to the generalised polar angle $i_{\rm s}$.   

\cite{Cuello2019} reported dynamical maps for polar orbits for massive planets with chaotic regions. They also find that polar circumbinary planets around eccentric binaries are less stable for low binary mass ratios (see their Figure 8).
\citet{Giuppone2019} investigated the orbital stability of a planet with $i=90^\circ$ orbiting a binary with $e_{\rm b}$ = 0.5 for various values of $f_{\rm b}$ and $m_{\rm p}$ (see e.g., their figure 2). Their results for $m_{\rm p}$ = 0.001 $\rm M_{\odot}$ shows that the planet is less stable for low binary mass ratios and they are consistent with our simulation results in the middle-left panel of Fig.~\ref{fig:map2}.

\begin{figure}
    \centering
    \includegraphics[width=8.7cm]{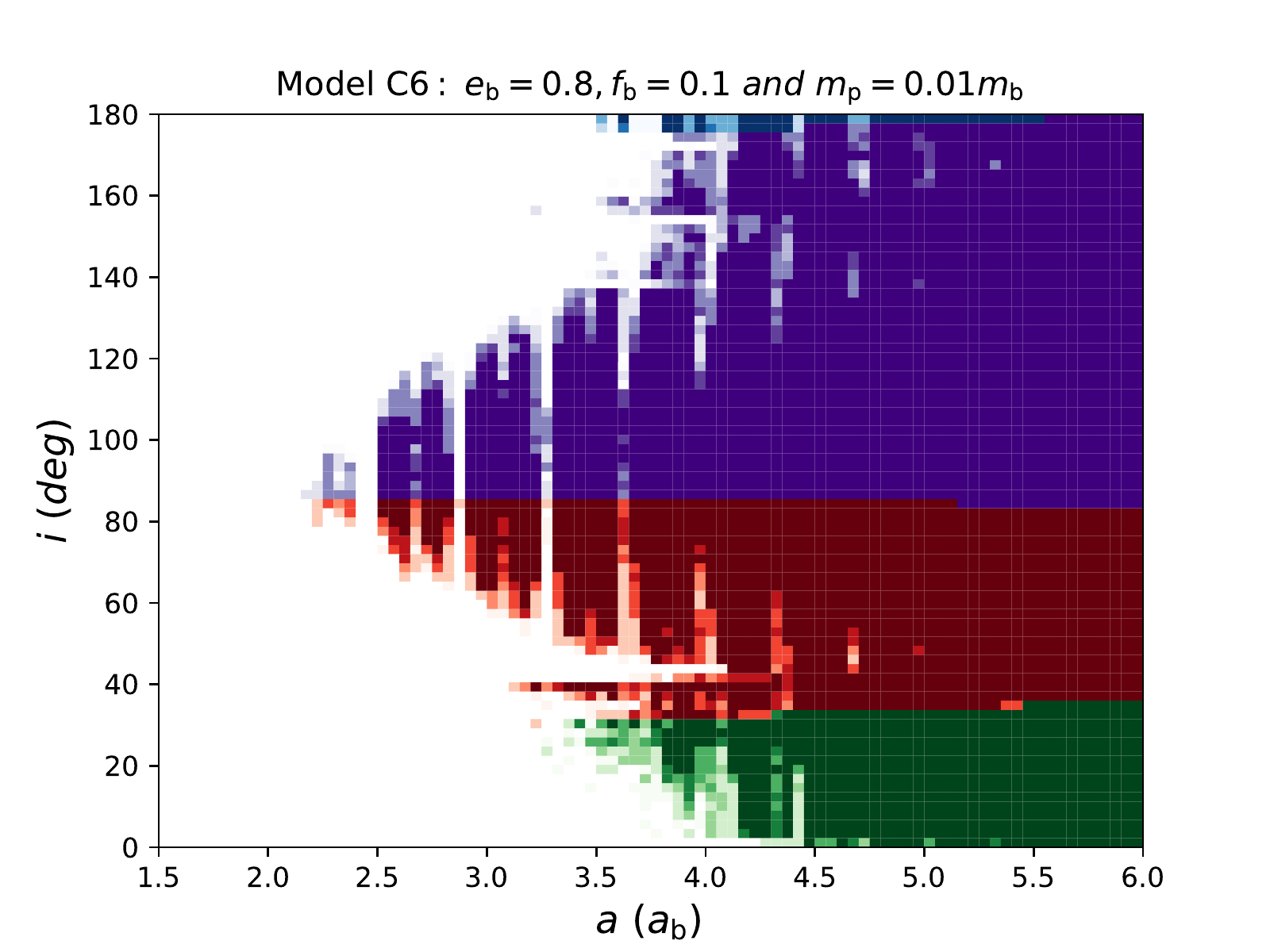}
    \caption{Same as Model C6 in Fig.~\ref{fig:map2} with the high accuracy IAS15 integrator and based on the MEGNO indicator. The colour points are for MEGNO indicator \textit{<Y>} between 1.5 and 2.4 at a time of $5 \times 10^4 T_{\rm b}$ that suggests orbital stability.}
    \label{fig:megno}
\end{figure}

As a check on our results, we applied  the Mean Exponential Growth of Nearby Orbits  (MEGNO) indicator value \textit{<Y>} that can identify whether the orbit is chaotic or quasi-stable  \citep{Cincotta2000}. If \textit{<Y>} converges to a value 2 over time, the orbit is stable, while if \textit{<Y>}  diverges linearly in time, the orbit is chaotic. To test how the MEGNO indicator compares with our results, we ran the most extremely eccentric and high mass case (Model C6) shown in the lower right of Fig.~\ref{fig:map2} using the high accuracy IAS15 integrator. The resulting stability map in Fig.~\ref{fig:megno} shows most of the stable orbits in the lower right of Fig.~\ref{fig:map2} have \textit{<Y>} $\simeq$ 2. Therefore, the MEGNO results are quite similar and support our conclusion that the generalised polar orbits are the most stable for high binary eccentricity.

\subsection{Stability maps for initial $\phi=0^\circ$}

In the case with initial $\phi=0^\circ$ there are no librating orbits. This can be seen for example from Fig.~1 in \cite{Chen20192}. Along the horizontal line in the $i\cos \phi - i \sin \phi$ phase plot there are no librating orbits.

Fig.~\ref{fig:phi=0fb=5} shows the stability maps for initial $\phi=0^\circ$ and an equal mass binary. Unlike Fig.~\ref{fig:map1} for the $\phi=90^\circ$ case, we see that the librating region is absent. The circulating regions are enlarged with similar features. Initial inclination of $90^\circ$ means that the initial angular momentum vector of the planet orbit is perpendicular to both the binary angular momentum vector and the binary eccentricity vector, in the direction of the $\bm{j}\times \bm{e_{\rm b}}$ vector. 

Fig.~\ref{fig:phi=0} shows the stability maps for initial $\phi=0^\circ$ and binary mass fraction $f_{\rm b}=0.1$. The orbits are generally less stable compared to the equal mass binary case. There is a more sensitive dependence on the mass of the planet in this case, as we found for the initial $\phi=90^\circ$ case.  The stability of the orbits in the $\phi=0^\circ$ case are quite similar to the stability for the circulating orbits in the $\phi=90^\circ$ case, but extend to a wider range of initial inclinations. 

The critical initial inclination above which the orbits are circulating and retrograde  (inclination at the green--blue boundary) increases with increasing planet mass. For the equal mass binary simulations, the critical inclination  of the retrograde orbits with a high mass planet is about 100$^\circ$. For small binary mass fraction, the critical initial inclination is larger than in the corresponding equal mass binary case and increases with increasing binary eccentricity. The critical initial inclination with high planet mass and low binary fraction increases to about 140$^\circ$ for $e_{\rm b} = 0.8$.

\subsection{Circumbinary planet escape time}

\begin{figure*}
    \centering
    \includegraphics[width=18cm]{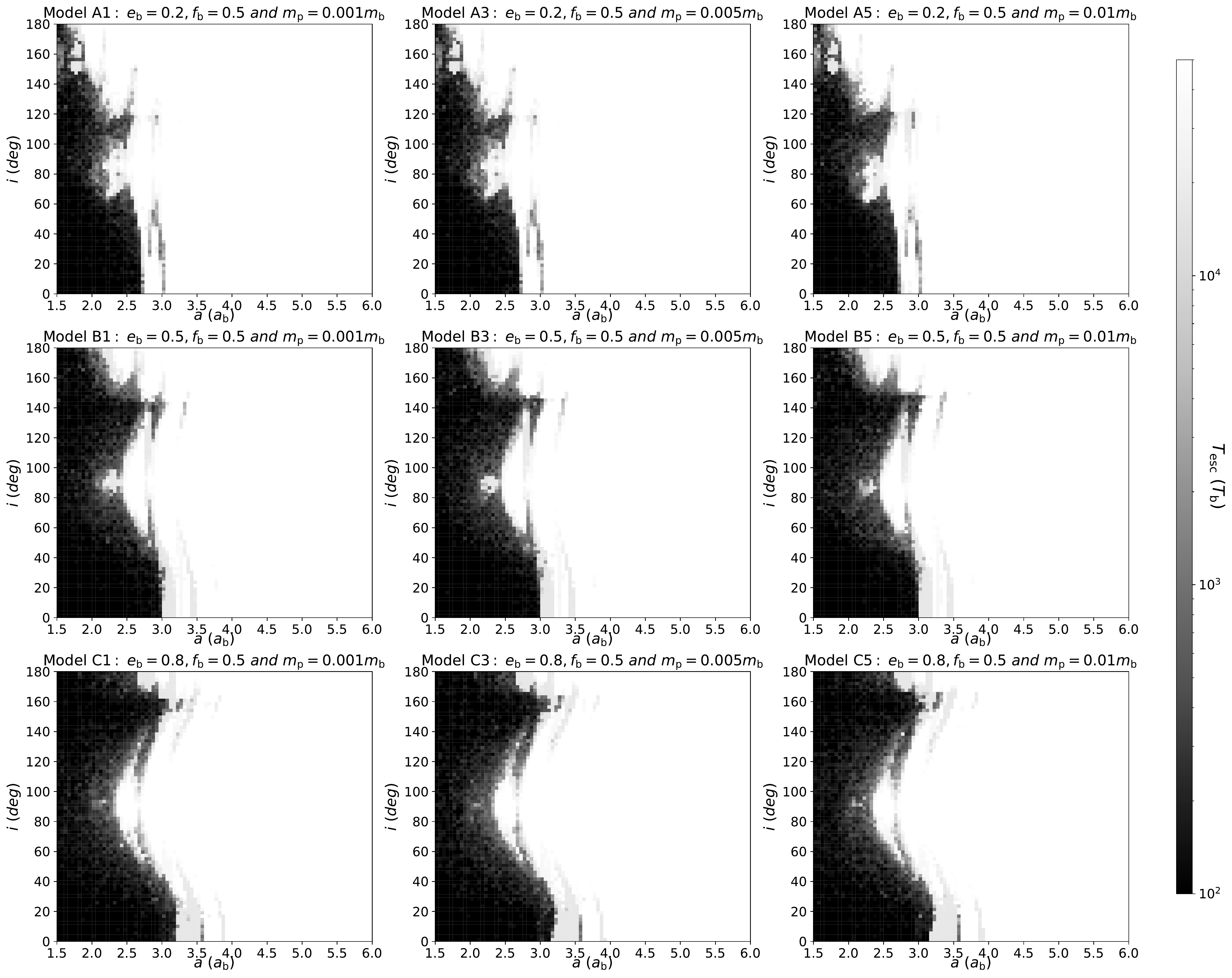}
    \caption{Average escape time, $t_{esc}$, as a function of planet semi--major axis $a$ for a binary system with binary mass fraction $f_{\rm b}$=0.5 and initial binary eccentricity $e_{\rm b}$ = 0.2 (first row), 0.5 (second row), and 0.8 (third row). The third body has mass  $m_{\rm p}=10^{-3}\,m_{\rm b}$ (first column), $5\times 10^{-3}\,m_{\rm b}$ (second column), and $10^{-2}\, m_{\rm b}$ (third column).}
    \label{fig:map3}
\end{figure*}

\begin{figure*}
    \centering
    \includegraphics[width=18cm]{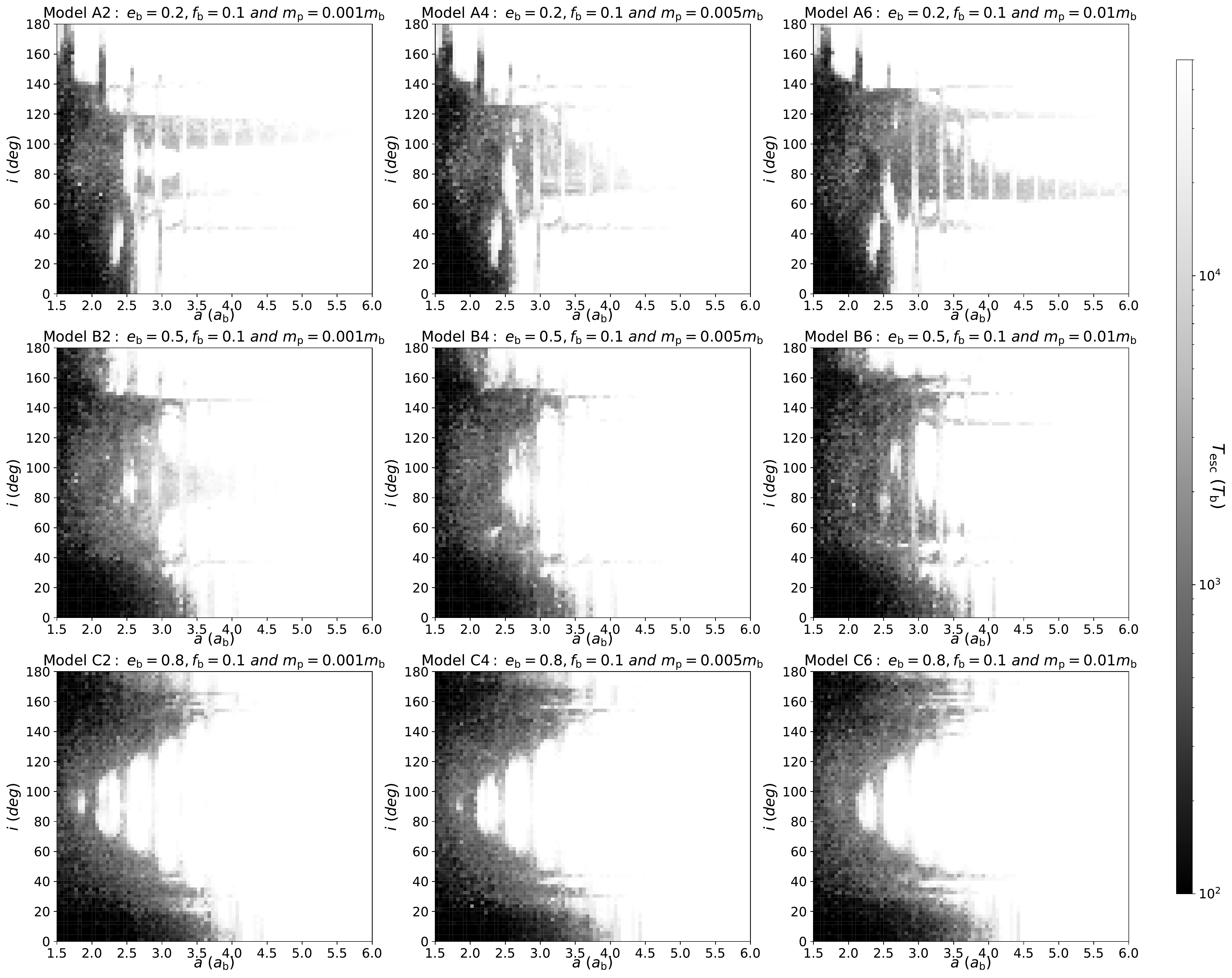}
    \caption{Same as Fig~\ref{fig:map3} except the binary mass fraction is $f_{\rm b}$=0.1.}
    \label{fig:map4}
\end{figure*}

Figs.~\ref{fig:map3} and~\ref{fig:map4} show the escape times of unstable orbits, $t_{\rm esc}$. The panels in Figs.~\ref{fig:map3} and~\ref{fig:map4} correspond to the panels in Figs.~\ref{fig:map1} and~\ref{fig:map2}, respectively. Each pixel in the density plot represents the average $t_{\rm esc}$ of a orbit. Darker pixels correspond to shorter escape times and a white pixel represents a stable orbit. 

We find that a system with smaller $f_{\rm b}$ has more long-lived unstable orbits (more light pixels) and a system with equal mass fraction models has more short-lived unstable orbits (more dark pixels), similar to that described in \citet{Doolin2011} for the test particle orbits.

Moreover, orbits with lower initial inclination are generally more short-lived and the number of short-lived unstable orbits with high initial  inclination increases with increasing binary eccentricity. Orbits with  initial inclination close to the stationary inclination, are more long-lived orbits even in the innermost region especially for models with the small $f_{\rm b}$. For more extreme mass ratio binaries, the binary torques due to resonances are weaker and therefore the unstable planet orbits evolve more slowly.

Finally, we do not observe a significant difference for unstable orbits between systems with different planet mass in the equal mass binary models. On the other hand,  there are more long-lived unstable orbits with increasing planet mass in the lower $f_{\rm b}$ models especially for $e_{\rm b} =0.2$.

\section{Discussion and Conclusions}
\label{conclusion}

In this paper, we have investigated the orbital stability of a misaligned initially circular orbit close-in planet with nonzero mass around an eccentric orbit binary by means of numerical simulations. In our suite of simulations, we have considered  planets with masses $m_{\rm p}$ = 0.001, 0.005 and $0.01\, m_{\rm b}$ around a binary with mass fraction $f_{\rm b} = 0.1$ and 0.5 and we sample different orbits by varying the initial semi-major axis, $a$,  inclination, $i$ and true anomaly, $\nu$.  We consider two values for the initial nodal phase angle, $\phi=0^\circ$ and $\phi=90^\circ$. 

In general, we find that circumbinary planet orbits are stable over times of $5 \times 10^4 T_{\rm b}$ for initial orbital radii $a$ that are greater than about $2.2 a_{\rm b}$ to about $4 a_{\rm b}$, although there are cases where the instability extends to larger radii. The values depend on the planet and binary parameters. 

The results show that at high binary eccentricity, planet orbits near the generalised polar orbits are the most stable type of planet orbit. In particular, they are more stable than prograde and retograde coplanar orbits. The enhanced stability of these generalised polar orbits over coplanar orbits is most apparent for more a extreme mass ratio, high eccentricity binary, as seen in the bottom right of Fig.~\ref{fig:map2}.

The range of radii covered by stable nearly polar orbits increases with binary eccentricity. For high binary eccentricity, the resonant binary Lindblad torque on a low mass polar gas disc decreases with increasing binary eccentricity \cite[e.g.][]{Miranda2015,Lubow2018,Franchini2019b}. As a result, the disc inner edge of a polar disc extends closer to the binary centre of mass than it does for a prograde coplanar disc. Analogous results appear in the planet orbit case in that the radii for stable orbits extend closer to the binary centre of mass for polar orbits than is the case for coplanar orbits.  For the polar case involving a high eccentricity binary, the potential due to the binary is at each instant of time is nearly axisymmetric in the plane of the planet orbit. Consequently the torque exerted on the planet perpendicular to its orbit plane is relatively weak compared to a lower eccentricity case. Just the opposite behaviour would be expected for a coplanar planet. That is, with increasing binary eccentricity the binary potential becomes more nonaxisymmetric in the plane of the planet orbit leading to a larger torque. Such considerations support the simulation results we find that the polar orbits are more stable than coplanar ones.

The orbital alignment of circumbinary planets depends upon the final alignment of the circumbinary disc. A low mass disc may evolve towards either coplanar alignment or polar alignment at $i=90^\circ$. However, high mass discs may align to the generalised polar state at tilts that are less than $90^\circ$ \citep{MartinandLubow2019}. Assuming the disc evolves through instantaneously stationary (generalised polar) configurations, the disc may move towards a tilt of $90 ^\circ$ as it loses mass. Therefore, the timescale of the disc dispersal plays a role in the final inclination of a debris disc or protoplanet. Disc-planet interactions must also be taken into account.  A giant planet that opens a gap may not remain coplanar with a misaligned disc \citep{Lubow2016,Pierens2018,Franchini2019c}. Thus, a massive planet could undergo libration oscillations in its orbit even after the disc has dispersed.

Our results have implications on the possible orbits in which circumbinary planets can reside. The results in Figs.~\ref {fig:map1} and \ref{fig:map2}
suggest that for highly eccentric orbit binaries, planets found inward of about $3.5 a_{\rm b}$ can reside on only misaligned librating orbits including polar orbits. Planets found inward of about $2.5 a_{\rm b}$ around low
eccentricity binaries can reside on only circulating retrograde orbits. Planets at larger separations $\ga 4 a_{\rm b}$
can be stable for both librating and circulating orbits over a wide range of planet orbit inclinations. That is, orbital stability considerations alone do not provide a constraint
on the possible planet orbits, if the planet is far enough from the binary.
But based on disc simulations, we expect there to be a preference for highly inclined and polar planets around high eccentricity binaries \citep[e.g.,][]{MartinandLubow2018b}.
Whether a disc evolves to polar depends on its initial inclination and nodal phase.
The detailed occurrence frequency of polar disc configurations and therefore polar planets then depends on the initial disc
misalignment distribution that is not known.
But binaries with inclined circumbinary discs are found to occur for binary orbital periods longer than about 30 days  \citep{Czekala2019}.

 A promising technique for finding polar orbit planets is to use binary
eclipse timing variations (ETVs).
To date, planet detections using ETVs based on Kepler data have been reported for the inclined planet KIC 5095269b \citep{Getley2017, Borkovits2016}. Two additional systems KIC 07177553 and KIC 7821010 may also have an inclined planetary-mass object orbiting around a binary \citep{Borkovits2016}. Since the ETV technique is becoming well developed, it is likely that additional misaligned and polar circumbinary planets will be found by the future observations with TESS or PLATO \citep{Zhang2019}. We believe that more inclined planets will be found, especially around longer period, eccentric orbit binaries.

\section*{Acknowledgements}
C.C. acknowledges support from a UNLV graduate assistantship. We acknowledge support from NASA through grants NNX17AB96G and 80NSSC19K0443.   Computer support was provided by UNLV's National Supercomputing Center and simulations in this paper made use of the \textsc{rebound} code which can be downloaded freely at http://github.com/hannorein/rebound.

\bibliographystyle{mnras} 
\bibliography{main}
\bsp	
\label{lastpage}
\end{document}